\documentclass{article}

\usepackage[utf8]{inputenc}  
\usepackage[T1]{fontenc} 
\usepackage{dcolumn}
\usepackage{fullpage}
\usepackage{parskip}
\usepackage{amssymb}
\usepackage{amssymb,amsmath}
\usepackage{color}
\usepackage[usenames,dvipsnames,svgnames,table]{xcolor}
\usepackage[colorlinks,urlcolor=blue]{hyperref}
\usepackage{cite,color,url}
\usepackage{array}
\usepackage{longtable}
\usepackage{multirow}
\usepackage{graphicx}
\usepackage{caption}
\usepackage{float}
\usepackage{colordvi}
\usepackage{epsfig,psfrag,rotating,soul}
\usepackage{tabularx}
\usepackage{hyperref}

\usepackage{slashed}

\newcolumntype{M}{>{$\displaystyle}c<{$}}
\newcolumntype{L}{>{$\displaystyle}l<{$}}

\newcommand\Vtextvisiblespace[1][.3em]
{%
	\mbox{\kern.06em\vrule height.3ex}%
	\vbox{\hrule width#1}%
	\hbox{\vrule height.3ex}
}

\newcommand{\beq}{\begin{eqnarray}} 
	\newcommand{\eeq}{\end{eqnarray}}

\newcommand{\hmssm}{\ensuremath{h\mathrm{MSSM} \ }}
\newcommand{\mhmodp}{\ensuremath{m_{h}^{\mathrm{mod}+} \ }}

\begin{document}

\vspace{1cm}

\begin{center}
\Large\bf\boldmath
\textcolor{black}{Signal versus Background Interference in $H^+\to t\bar b$ Signals \\
for MSSM Benchmark Scenarios}
\unboldmath
\end{center}
\vspace{0.05cm}
\begin{center}
Abdesslam Arhrib$^a$,
Duarte Azevedo$^b$,
Rachid Benbrik$^c$,
Hicham Harouiz$^{c}$,
Stefano Moretti$^d$,\\[0.2cm]
Riley Patrick$^{e}$,
Rui Santos$^{b,f}$
\\[0.4cm]
{\small
{\sl${}^a$Faculty of Sciences and Techniques, Abdelmalek Essaadi University, B.P. 416, Tanger, Morocco}\\[0.2em]  
{\sl${}^b$Centro de F\'{\i}sica Te\'{o}rica e Computacional,     Faculdade de Ci\^{e}ncias, Universidade de Lisboa, Campo Grande, Edif\'{\i}cio C8,
  1749-016 Lisboa, Portugal}\\[0.2em]
{\sl${}^c$MSISM Team, Facult\'e Polydisciplinaire de Safi, Sidi Bouzid, B.P. 4162, Safi, Morocco}\\[0.2em]
{\sl${}^d$School of Physics and Astronomy, University of Southampton,
Southampton, SO17 1BJ, United Kingdom}\\[0.2em]
{\sl${}^e$ARC Center of Excellence for Particle Physics at the Terascale, Department of Physics, University of Adelaide, 5005 Adelaide, South
Australia}\\[0.2em]
{\sl${}^f$ISEL - Instituto Superior de Engenharia de Lisboa, Instituto Polit\'ecnico de Lisboa,  1959-007 Lisboa, Portugal}\\[0.2em]
}

\end{center}
\vspace*{1mm}

\vspace*{1.0truecm}
\begin{center}
{\bf \large Abstract}
\end{center}

\noindent
In this paper, we investigate sizeable  interference effects between a {heavy}  charged Higgs boson signal produced dominantly via  $gg\to t\bar b H^-$ (+ c.c.) followed by the decay  $H^-\to b\bar t$ (+ c.c.)  and the irreducible background given by $pp\to t\bar t b \bar b$ topologies at the Large Hadron Collider (LHC). We show that it may be possible that such effects could spoil current  $H^\pm$ searches where signal and background  are normally treated separately. The reason  for this is that a heavy charged Higgs boson can have a large total width, in turn enabling such interferences, altogether leading to potentially very significant alterations, both at the inclusive and exclusive level, of the yield induced by the signal alone.  This therefore implies that currently established LHC searches for such wide charged Higgs bosons might require modifications. We show such effects quantitatively using two different benchmark configurations of the minimal realisation of Supersymmetry, wherein such $H^\pm$ states naturally exist. However, on the basis of the limited computing resources available, we are unable to {\sl always} bring the statistical error down to a level where all such interference effects are unequivocal, so that we advocate dedicated experimental analyses to confirm this with higher statistics data samples.   
 
\clearpage


\section{Introduction}
After the discovery of a Higgs-like particle  at Large Hadron Colliders (LHC) a few years ago \cite{Aad:2012tfa,Chatrchyan:2012xdj}, a significant amount of both theoretical and experimental activities have taken place trying to identify the nature of this object.
The mass of such a particle and its couplings to some Standard Model (SM) particles are now measured with a good precision 
\cite{Aad:2015zhl,Khachatryan:2016vau}. Their values indicate that such a Higgs-like particle is light and its properties (spin, CP quantum numbers and interactions) are consistent with those of the SM Higgs boson.

However, there are many theoretical and experimental  indications that show that the SM can only be an effective theory of a more fundamental one that still needs to be discovered.  Many Beyond the SM (BSM) scenarios have been put forward over the years and it is fair to say that one stems as the most appealing one -  Supersymmetry (SUSY). This is because it solves the well-known hierarchy problem of the SM by protecting the  Higgs mass from unstable higher order corrections thanks to the new symmetry between fermions and bosons that it predicts \cite{Book}. Furthermore, SUSY also has the capability to address the Dark Matter (DM) and gauge unification problems of the SM, indeed,  without any proliferation of fundamental parameters if one assumes that SUSY can in turn be  viewed as an effective realisation  of some Grand Unified Theory (GUT), like Supergravity \cite{Nilles:1983ge,Haber:1984rc}. The Minimal Supersymmetric Standard Model (MSSM) is the simplest realisation of SUSY that predicts a light Higgs boson $h^0$ that can be identified as the observed 125 GeV Higgs-like particle and can be as successful as the SM when confronted with experimental data, yet it can surpass it in all the above respects. 

The Superpotential of the MSSM has to be holomorphic, thus one needs to introduce at least two Higgs doublets fields, one more than in the SM. One of these generates masses for up quarks and the other one generate masses for down quarks and charged leptons.
From the 8 degrees of freedom present in such a 2-Higgs Doublet Model (2HDM), 3 are acquired by the longitudinal components of the gauge bosons $W^\pm$ and $Z^0$, so that the latter get a mass too, and the remaining 5 appears as new Higgs particles$:$ 2 CP-even $h$ and $H$ (with $M_{h^0}<M_{H^0}$), 
one CP-odd $A$ and a pair of charged ones $H^\pm$. 
Discovery of any such new states would be unmistakable evidence of BSM physics, yet, only charged Higgs states would be a clear hint towards a 2HDM structure, as required by the MSSM, as additional neutral Higgs states could be attributed to singlet structures entering an extended Higgs sector. 

At a hadron collider, production of charged Higgs bosons proceeds through many channels. If the charged Higgs boson is light (i.e., $M_{H^\pm}<m_t+m_b$), 
it can be produced from top or anti-top decay$:$ i.e., 
$gg,q\bar q\to t\bar{t}$ followed by, e.g.,  $\bar t \to \bar{b}H^-$ (+ c.c.). Given the fact that the $t\bar{t}$ cross section is very large, 
this mechanism would give an important source of light charged Higgs states.
After the top-bottom threshold (i.e., $M_{H^\pm}>m_t+m_b$), a charged Higgs boson can be produced in association with top-bottom pairs, i.e., $bg\to t H^- $ \cite{single}. In fact, these two channels are captured at once
by the  $gg\to  t\bar{b}H^-$ (+ c.c.) `complete' process, as explained in \cite{Assamagan:2004gv,Guchait:2001pi}.
There  exist other production  mechanisms too, such as $pp\to H^+H^-$, 
$pp\to A^0H^\pm$,  $pp\to H^\pm W^\mp$, etc. which are however subleading compared to the previous  ones\footnote{For a recent review, see\cite{Akeroyd:2016ymd}.}.

At the LHC, a light charged Higgs boson, with $M_{H^\pm} < m_t +m_b$, can be detected from $t\bar{t}$ production 
followed by top or anti-top quark decay $t\to b H^+$ if the $H^-$ state 
decays dominantly to  $\tau\nu$. ATLAS and CMS have already set a limit on  ${\rm BR}(t\to bH^+)\times {\rm BR}(H^+\to \tau\nu)$
\cite{Aad:2014kga, Khachatryan:2015qxa,Sirunyan:2019hkq,Aaboud:2018gjj},  which can be translated into a limit on the $(M_{H^\pm},\tan\beta)$ plane, where $\tan\beta$ is the ratio of the two Higgs doublet 
vaccuum expectation values (VEVs). {In the MSSM, for some specific benchmark scenarios, charged Higgs bosons with mass less than about 160 GeV are ruled out for almost any value of $\tan\beta$ 
\cite{Sirunyan:2019hkq,Aaboud:2018gjj}.}
However, heavy charged Higgs states, with $M_{H^\pm} > m_t +m_b$, are generally allowed as they would decay dominantly 
into $t\bar{b}$, which is a rather difficult final state to extract due to large reducible and irreducible backgrounds associated with jets emerging from 
$H^-\to \bar{t}b$ decays. Even then, one could still get a moderate signal from such a channel for small $\tan\beta\leq 1.5$ or large $\tan\beta\geq 40$ 
\cite{Miller:1999bm,Moretti:1999bw}. Another possibility for detecting heavy charged Higgs states would be the search for $H^+\to \bar{\tau}\nu_\tau$ (i.e., like the preferred one for a light state), which enjoys a smaller background
in comparison. At the LHC Run 2, both channels have been searched for and no excess over the background only hypothesis have been reported.
Therefore, limits are set on $\sigma(pp\to tH^-)\times {\rm BR}(H^- \to \bar{t}b/\tau\bar{\nu}_\tau)$ (+ c.c.) \cite{Aaboud:2016dig,CMS:2016szv,ATLAS:2016qiq,Aaboud:2018cwk}.
In the MSSM, one can have additional SUSY channels  that can contribute to $H^\pm$ production and/or decay, e.g.,  production from squark/gluino cascades \cite{Moortgat:2001pp,Bisset:2005rn} and/or 
decays  into chargino-neutralino states  \cite{Bisset:2000ud,Bisset:2003ix}, though these require special MSSM configuration assumptions, hence they are not currently pursued by ATLAS and CMS.

The current highest priority, in relation to charged Higgs boson searches at the LHC, is to further establish the $H^+\to t\bar b$ decay channel in the
heavy mass region.  With this in mind, using the framework of a generic 2HDM~\cite{Arhrib:2017veb}, 
we have investigated the possibility of having large interference effects 
between signals from a heavy charged Higgs boson via $bg\to tH^-\to  t W^- A\to t W^-b\bar{b}$ (and similarly for $h$ and $H$) 
and the irreducible background from $bg\to tW^-b\bar{b}$ processes. Therein, it was shown that such interference effects can  
modify any dedicated  charged Higgs boson searches where signal and background are treated separately, which is the case for all aforementioned experimental 
analyses.

The purpose of this paper is to address similar  issues for the MSSM, i.e., to quantify the impact of interference effects between the `complete' signal 
$pp \to t\bar b H^- $ and the irreducible background in the $H^-\to b\bar t$ decay channel. 
We will show that such effects are indeed large for heavy $H^\pm$ masses for two MSSM benchmark scenarios, both for inclusive cross section calculations and after a full detector  analysis. The plan of the paper is as follows. In the forthcoming section we describe the  MSSM configurations used. Sect. 3 dwells on the MSSM spectra conducive to generate such large interference phenomena. Sect. 4 presents our numerical results. Finally, we conclude in Sect. 5.

\section{Definition of the benchmark scenarios}
At tree level, the MSSM Higgs sector is completely fixed by 2 parameters$:$ $\tan\beta$ and a Higgs boson mass, e.g., the CP-odd one ($M_{A^0}$).
One of the major predictions of SUSY is the presence of a light CP-even Higgs (lighter than $Z$ boson at the lowest order) in the spectrum. {However, high order corrections can shift such a  mass 
in order to fit the observed Higgs-like particle mass \cite{Heinemeyer:1998np,Heinemeyer:1998jw,Heinemeyer:1998kz}.  It has been shown in \cite{Degrassi:2002fi}
that high order corrections could raise the tree-level MSSM prediction for such a mass up to 135 GeV for large soft trilinear 
breaking terms and also that the theoretical uncertainties due 
to the unknown high order effects should be of the order of 3 GeV}.


In the MSSM, the most important parameters  relevant for the prediction of the 
masses, couplings and, hence, production cross sections and decay 
probabilities of the Higgs bosons are$:$  
$\tan\beta$, $M_{A^0}$, the soft SUSY-breaking masses for the stop and sbottom squarks 
(which, for simplicity, we assume all equal to a common mass parameter $M_S$), the soft
SUSY-breaking gluino mass $m_{\tilde {g}}$, 
the Superpotential Higgs-mass parameter $\mu$ and the left-right
mixing terms in the stop and sbottom mass matrices, i.e.,   
\begin{eqnarray}
X_t = A_t - \mu \cot\beta,\qquad \qquad X_b = A_b - \mu\tan\beta,
\label{eq_def_Xt}
\end{eqnarray}
 respectively.  
 
{We  use the PROSPINO public code~\cite{Beenakker:1996ed} to compute the  
charged Higgs boson production cross section $\sigma(pp\to t (\bar b)H^-  + {\rm c.c.})$, 
which includes Next-to-Leading Order (NLO) corrections to the $bg\to tH^-$ + c.c.  (2-to-2) process. We use the inclusive cross section computed this way  to test the viability of our proposed MSSM scenarios against data. 

However, we adopt the tree-level $pp\to t\bar b H^-$ + c.c. (2-to-3) process for Monte Carlo (MC) event generation, because it produces a better description of the signal at the differential level in the detector region than the former channel (i.e., the additional $b$-(anti)quark is explicit in the phase space rather than integrated into the proton content) and because the corresponding irreducible background is only known at LO. This clearly implies that the normalisation used for the MC analysis is different from that used in the inclusive parameter scans, however, we note that we are primarily concerned here with the relative behaviour of signal, irreducible background and relative interference, rather than the overall normalisation. (Note that, hereafter, we always sum over both $H^+$ and $H^-$.)}  Both PROSPINO and FeynHiggs  \cite{Heinemeyer:1998yj,Hahn:2009zz} use the same (on-shell) renormalisation scheme, therefore, 
 the input values of the MSSM parameters can be passed seamlessly from the Higgs spectrum generator   
  to the cross section calculator. 
  The MSSM parameter space is already highly constrained by asking that one of
the CP-even neutral scalar states should mimic the properties of the SM-like Higgs 
boson observed at LHC while the additional Higgs bosons should satisfy the existing constraints obtained 
from ATLAS and CMS from different channels. {For this purpose,  the FeynHiggs code is  linked to {HiggsBounds-5.2.0beta}~\cite{Bechtle:2008jh,
  Bechtle:2011sb,Bechtle:2013wla,Bechtle:2015pma} and 
  {HiggsSignals-2.2.0beta}~\cite{Bechtle:2013xfa}  
 allowing us to check the consistency of our parameter space against various LHC as well as Tevatron and LEP constraints. We list  in Tab.~\ref{table1} the specific charged Higgs boson searches that have been included in HiggsBounds.}
 \begin{table}
    \scriptsize
    \begin{center}
        \begin{tabular}{|c|c|c|c|}
            \hline
            Experiment & Luminosity [fb$^{-1}$] & Label & Channel \\
            \hline
            LEP &          -- & \cite{LEPI} & $e^+e^- \to H^+H^- \to qq'qq'$ \\
            \hline
            LEP &          -- & \cite{LEPII} (DELPHI)& $e^+e^- \to H^+H^- \to qq'qq'$ \\
            \hline
            LEP &          -- & \cite{LEPII} (DELPHI) & $e^+e^- \to H^+H^- \to \tau\nu\tau\nu$ \\
            \hline
            D0  &       1.000 & \cite{Abazov:2009aa} (D0) & $t \to bH^+ \to bqq'$ \\
            \hline
            CDF &       2.200 & \cite{Aaltonen:2009ke} (CDF) & $t \to H^+ b$ \\
            \hline
            CDF &       0.192 & \cite{CDFNote7712} (CDF)& $t \to bH^+ \to b\tau\nu$ \\
            \hline
            CDF &       0.335 & \cite{CDFNote8353} (CDF) & $t \to bH^+ \to b\tau\nu$ \\
            \hline
            D0  &       1.000 & \cite{Abazov:2009aa} (D0) & $t \to bH^+ \to b\tau\nu$ \\
            \hline
            ATLAS, 7 TeV &    0.035 & \href{}{ATLAS-CONF-2011-094                                          } & $t \to H^+b \to c\bar{s}b$ \\
            \hline
            ATLAS, 7 TeV  &    4.600 & \cite{Aad:2012tj} & $t \to H^+ b$ \\
            \hline
            ATLAS, 8 TeV  &   19.500 & \href{}{ATLAS-CONF-2014-050                                          } & $t \to bH^+ \to b\tau\nu$ \\
            \hline
            ATLAS, 13 TeV &   36.100 & \cite{Aaboud:2018gjj} & $pp \to tbH^+ \to tb\tau\nu$ \\
            \hline
            ATLAS, 13 TeV &   36.100 & \cite{Aaboud:2018gjj} & $t \to bH^+ \to b\tau\nu$ \\
            \hline
            CMS , 8 TeV &   19.700 & \href{}{CMS-PAS-HIG-14-020                                           } & $t \to bH^+ \to b\tau\nu$ \\
            \hline
            CMS, 8 TeV &   19.700 & \href{}{CMS-PAS-HIG-13-035                                           } & $t \to H^+b \to c\bar{s}b$ \\
            \hline
            CMS, 8 TeV &   19.700 & \href{}{CMS-PAS-HIG-16-030                                           } & $t \to H^+b \to c\bar{b}b$ \\
        \hline
            CMS, 13 TeV &   12.900 & \cite{CMS:2016szv} & $t \to bH^+ \to b\tau\nu$ \\
            \hline
            CMS, 13 TeV &   35.900 & \href{}{CMS-PAS-HIG-18-014                                           } & $t \to H^+b \to c\bar{b}b$ \\
            \hline
            ATLAS, 13 TeV &   14.700 & \href{}{ATLAS-CONF-2016-088                                          } & $pp \to tbH^+ \to tb\tau\nu$ \\
            \hline
            ATLAS, 13 TeV &   36.100 & \cite{Aaboud:2018cwk} & $pp \to tbH^+ \to ttbb$ \\
            \hline
        \end{tabular}
        \caption{Constraints on charged Higgs boson processes implemented in HiggsBounds and used in our analyses.}
           \label{table1}
    \end{center}
\end{table}
Additionally, a variety of lower energy constraints have been enforced, 
such as $B\to \tau \nu$, $B_{d,s}\to \mu^+ \mu^-$, $B\to X_s \gamma$ and $\Delta m_{s,d}$
(see details in \cite{Arhrib:2018ewj}).

All of the  MSSM benchmark scenarios adopted in our analysis  are characterised by relatively large values of the ratio $X_t/M_S$. This ensures that the mass of the SM-like Higgs state falls within the required 
range without the need for an extremely heavy stop.  In addition, the gaugino mass parameters, $M_2$ and $M_1$, are usually assumed to be related via the GUT relation
\begin{equation}
M_1 = \frac{5}{3}\frac{\sin^2\theta_W}{\cos^2\theta_W}M_2.
\end{equation}
We set the Higgs-sfermion interaction terms $A_f$ to zero for the first and second generation fermions$:$ $f=u,d,c,s,e,\mu$.
Moreover, the masses of the gluino and  first two generation squarks are set to 1.5 TeV, large enough to evade the current ATLAS and
CMS  limits from SUSY searches. 
\begin{table}[!t]
    \centering
    {\renewcommand{\arraystretch}{1} 
    {\setlength{\tabcolsep}{0.01cm} 
    \begin{tabular}{|c||c|c|c|c|c|}
        \hline
        MSSM Scenarios & \hmssm & \mhmodp \\
        \hline\hline
        $\tan\beta$ & 1--15 & 1--25   \\
        \hline
        $M_{A^0}$ (GeV) & 150--1000 & 90--1000  \\
        \hline
        $M_{{Q}_{1,2}}=M_{{U}_{1,2}}=M_{{D}_{1,2}}$ (TeV) & - & 1.5 \\
        $M_{{Q}_{3}}=M_{{U}_{3}}=M_{{D}_{3}}$ (TeV) & - & 1   \\
        \hline
        $M_{{L}_{1,2}}=M_{{E}_{1,2}}$ (TeV) & - & 0.5 \\
        $M_{{L}_3}=M_{{E}_3}$ (TeV) & - & 1  \\
        \hline
        $\mu$ (TeV) & - & 0.2  \\
        \hline

        $X_{t}$ (TeV) & - & 1.5 \\
        $A_{t}$ (TeV) & -& $X_t+\mu/\tan\beta$  \\
        $A_{b}$ (TeV) & -& $A_{t}$   \\
        $A_{\tau}$ (TeV) & - & $A_{t}$  \\
        \hline
        $M_1$ (TeV) & - & GUT relation \\
        $M_2$ (TeV) & - & 0.2 \\
        $M_3$ (TeV) & - & 1.5   \\
        \hline
    \end{tabular}}}
    \caption{MSSM input parameters for our two MSSM benchmark scenarios.}
    \label{tab:benchmark_scenarios}
\end{table}
In Tab.~\ref{tab:benchmark_scenarios} we list the MSSM parameters needed for the evaluation of the spectrum. 
We now move on to a detailed description of the MSSM benchmark scenarios to be used here, known as 
 $m_{h}^{\rm mod+}$ \cite{Carena:2013ytb}  and $h$MSSM \cite{Djouadi:2013uqa}.

\subsection{The $m^{{\rm mod}+}_{h}$ scenario}
The $m^{{\rm mod}+}_{h}$ scenario is a modification of the so-called 
maximal mixing scenario $m^{\rm max}_{h}$ \cite{Heinemeyer:1999zf} which was introduced
to maximize $M_{h^0}$ value by incorporating large high order effects and also
to give conservative limit on $\tan\beta$ during Higgs boson searches at LEP. 
This scenario predicts a CP-even Higgs $M_{h^0}$ slightly larger than the observed Higgs mass
 and that is why $m^{\rm max}_{h}$ scenario has been modified in order to 
accommodate the observed Higgs of $125$ GeV. The modification is performed by reducing the amount 
of scalar top mixing  such that the mass of the lightest Higgs state,  $M_{h^0}$,  is compatible with the  mass of the observed Higgs  boson within $\pm 3$ GeV as a theoretical uncertainty. 
When confronting $m^{{\rm mod}+}_{h}$   with the LHC data \cite{Carena:2013ytb}, there is a substantial region in $(M_{A^0},\tan\beta)$ plan with $\tan\beta>7$ for which the light 
CP-even Higgs is in a good agreement with the measured Higgs mass at the LHC. The SUSY inputs for this scenario  are given in the second column of Tab.~\ref{tab:benchmark_scenarios} and the spectrum is computed by the use of FeynHiggs code.

\subsection{The $h$MSSM scenario}
In $m^{{\rm mod}+}_{h}$, one needs to input $M_{A^0}$, $\tan\beta$ and other SUSY parameters 
in order to make a prediction for $M_{h^0}$ within the allowed range, $[122,128]$ GeV.
However, plenty of points on the $(M_{A^0},\tan\beta)$ plane would correspond 
to one value of $M_{h^0}$, the SM-like Higgs boson mass. In order to avoid such a situation,
the $h$MSSM was introduced \cite{Djouadi:2013uqa} in which $M_{h^0}$ was enforced to be approximately 125 GeV as well as the SUSY breaking scale $M_{\rm SUSY}\approx M_S$ fixed to be rather high, $\gg 1$ TeV, in order to 
explain the non-observation of any SUSY particle at colliders. A key assumption of the $h$MSSM
is to assume that radiative corrections to the diagonal mass of the heavy CP-even Higgs, $\Delta M_{22}$, 
are much larger than the ones to the light CP-even Higgs, $\Delta M_{11}$,  
and the mixing term between $h$ and $H$, $\Delta M_{12}$ \cite{Djouadi:2013uqa}, that is: 
$\Delta M_{22}\gg \Delta M_{11}, \Delta M_{12}$.  Therefore, $\Delta M_{22}$, which 
parameterises the SUSY effects,  is traded 
for the experimental value of $M_{h^0}$, $\tan\beta$ and $M_Z$.
 Therefore, the $h$MSSM setup describes the MSSM Higgs sector in terms of just $M_{A^0}$ and $\tan\beta$, exactly like the tree-level predictions, given the experimental knowledge of $M_Z$ and $M_{h^0}$.
The SUSY input parameters in this scenario are given in the first column of 
Tab.~\ref{tab:benchmark_scenarios} and the spectrum is computed by the 
 HDECAY code \cite{Djouadi:1997yw}.

%

\section{Higgs boson masses and Branching Ratios (BR)}
{In our analysis, we include the measured signal rates from the ATLAS and CMS Run 2 results via
{Higgs\-Signals-2.2.0beta}~\cite{Bechtle:2013xfa}  
which returns a $\chi^2$ value for the consistency between the model predicted signal rates and the corresponding measurements. 
We then determine the minimal  $\chi^2$ value over the scanned parameter space, $\chi_{\rm min}^2$, 
and keep as allowed the portion of it  that features a  
$\chi^2$ value within $\Delta \chi^2 \equiv  \chi^2 - \chi_{\rm min}^2$. 
For every benchmark scenario, we show the $\Delta\chi^2$ behavior, the best fit point,  
the charged Higgs total width, the typical BRs for charged Higgs 
decays into various final states and the charged Higgs production cross section. }

\subsection{The \hmssm case}
\begin{figure}[H]
\includegraphics[scale=0.45]{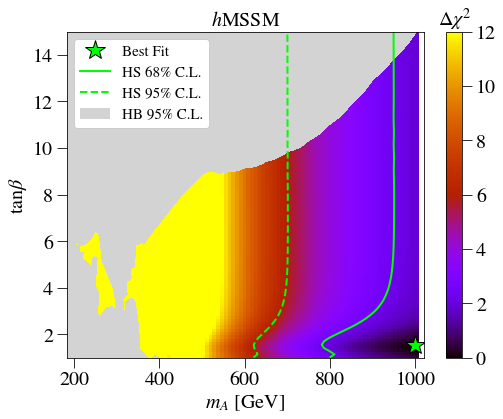}
\includegraphics[scale=0.45]{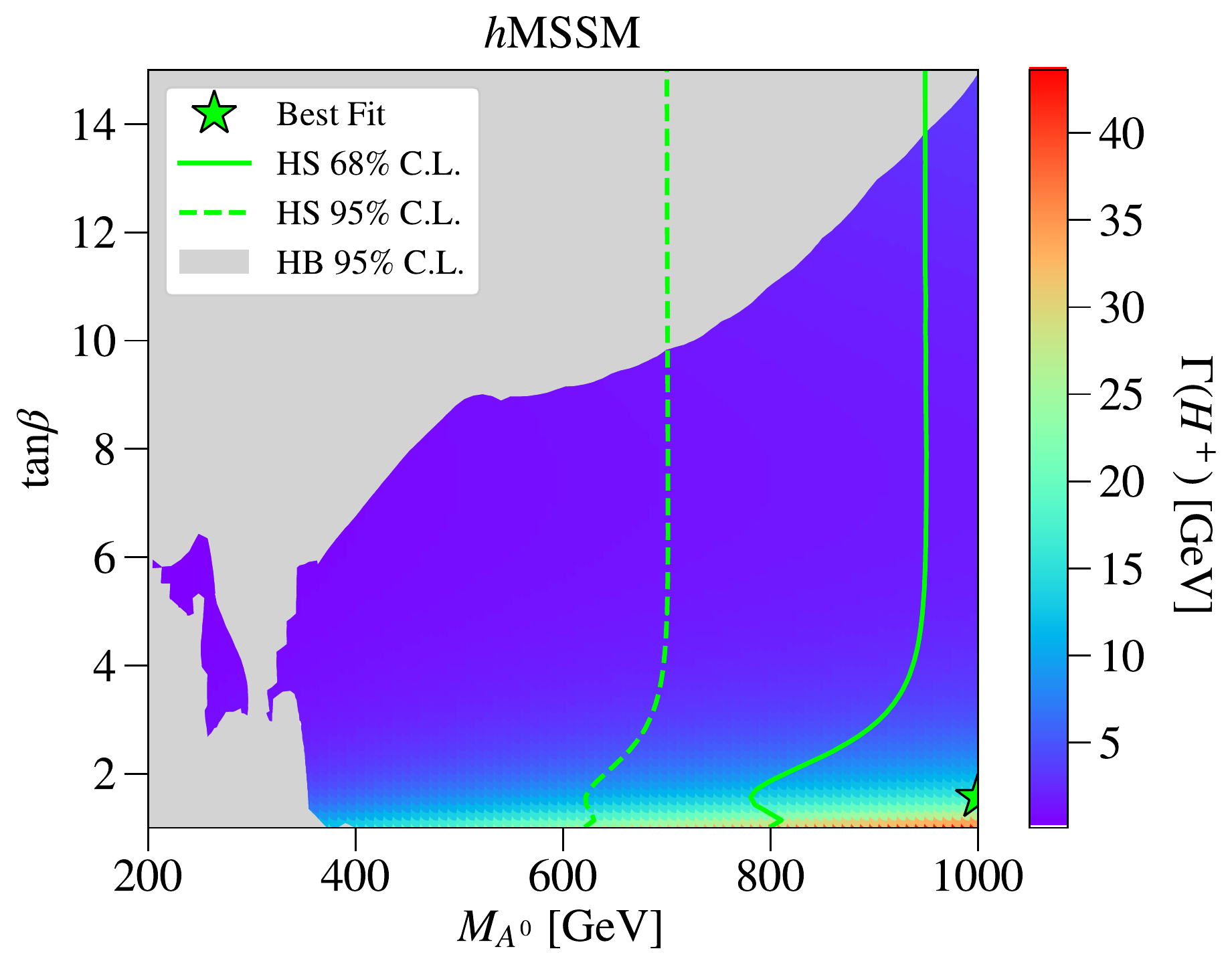}    
\includegraphics[scale=0.45]{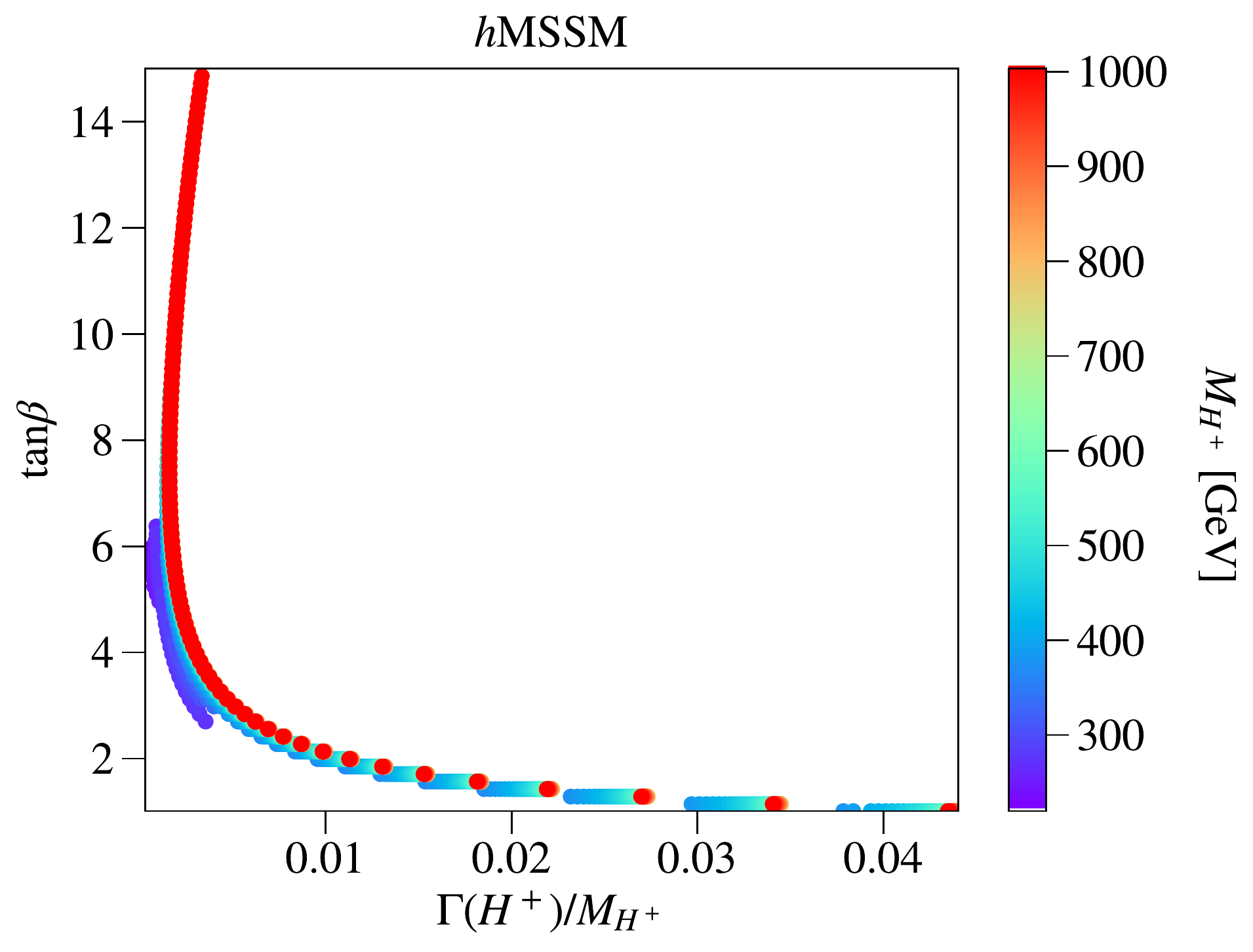}
\includegraphics[scale=0.45]{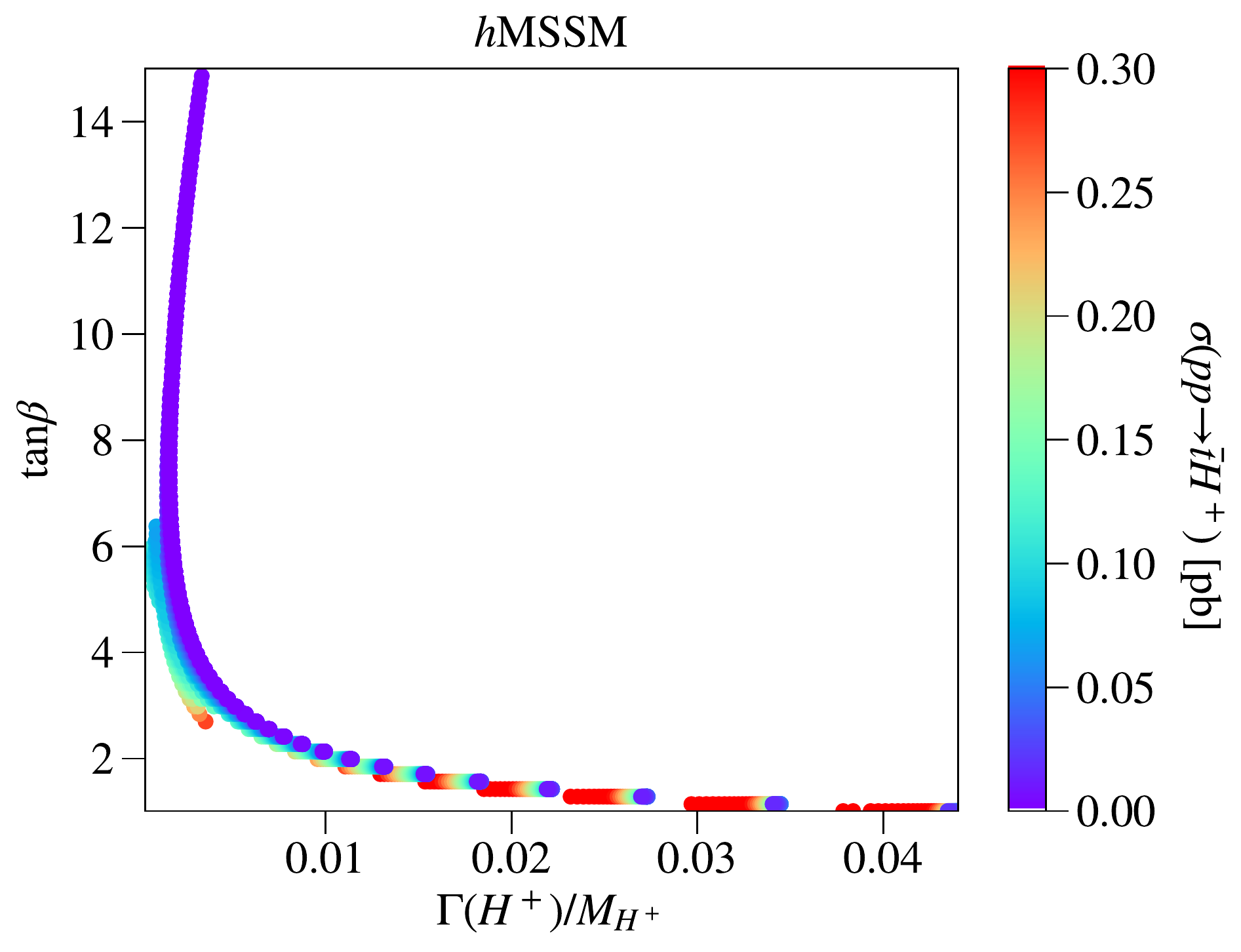}
\caption{
$\Delta\chi^2$ (top-left) and the charged Higgs total width (top-right) in the ($m_A\equiv M_{A^0}$, $\tan\beta$) plane. The best fit point is located at $M_{A^0}\approx 1$ TeV and $\tan\beta\approx 2$.  
The green lines show the exclusion limits from HiggSignals at $1\sigma$ (solid) and $2\sigma$ (dashed) while the gray 
area is ruled out by the various LHC searches implemented in HiggsBounds. The ratio $\Gamma_{H^\pm}/M_{H^\pm}$ as a function of the charged Higgs mass 
is shown in the bottom-left panel while in the bottom-right one it is presented as a function of the charged Higgs production cross section. }
\label{fig:hMSSM_HB5HS2_excl}
\end{figure}

In Fig.~\ref{fig:hMSSM_HB5HS2_excl}, we present $\Delta\chi^2$ (top-left) and the charged Higgs total width (top-right) in the ($M_{A^0}$, $\tan\beta$) plane. The best fit point is located at $M_{A^0}\approx 1$ TeV and $\tan\beta\approx 2$.  The green lines show the exclusion limits from HiggSignals at $1\sigma$ (solid) and $2\sigma$ (dashed) while the gray 
area is ruled out by the various LHC searches implemented in HiggsBounds. As one can see, the charged Higgs in the \hmssm scenario is rather heavy $\geq 550$ GeV 
and the total width is large for small $\tan\beta$ and gets reduced for high $\tan\beta$ values.
In the bottom panel we show the ratio $\Gamma_{H^\pm}/M_{H^\pm}$ as a function of the charged Higgs mass (left) as well as a function of the charged Higgs production cross section (right). The latter can be slightly above 1 pb. It is also visible from  the 
lower panel that the charged Higgs total width can be about 4\% of the charged Higgs mass at low $\tan\beta$.

\begin{figure}[H]
\includegraphics[scale=0.45]{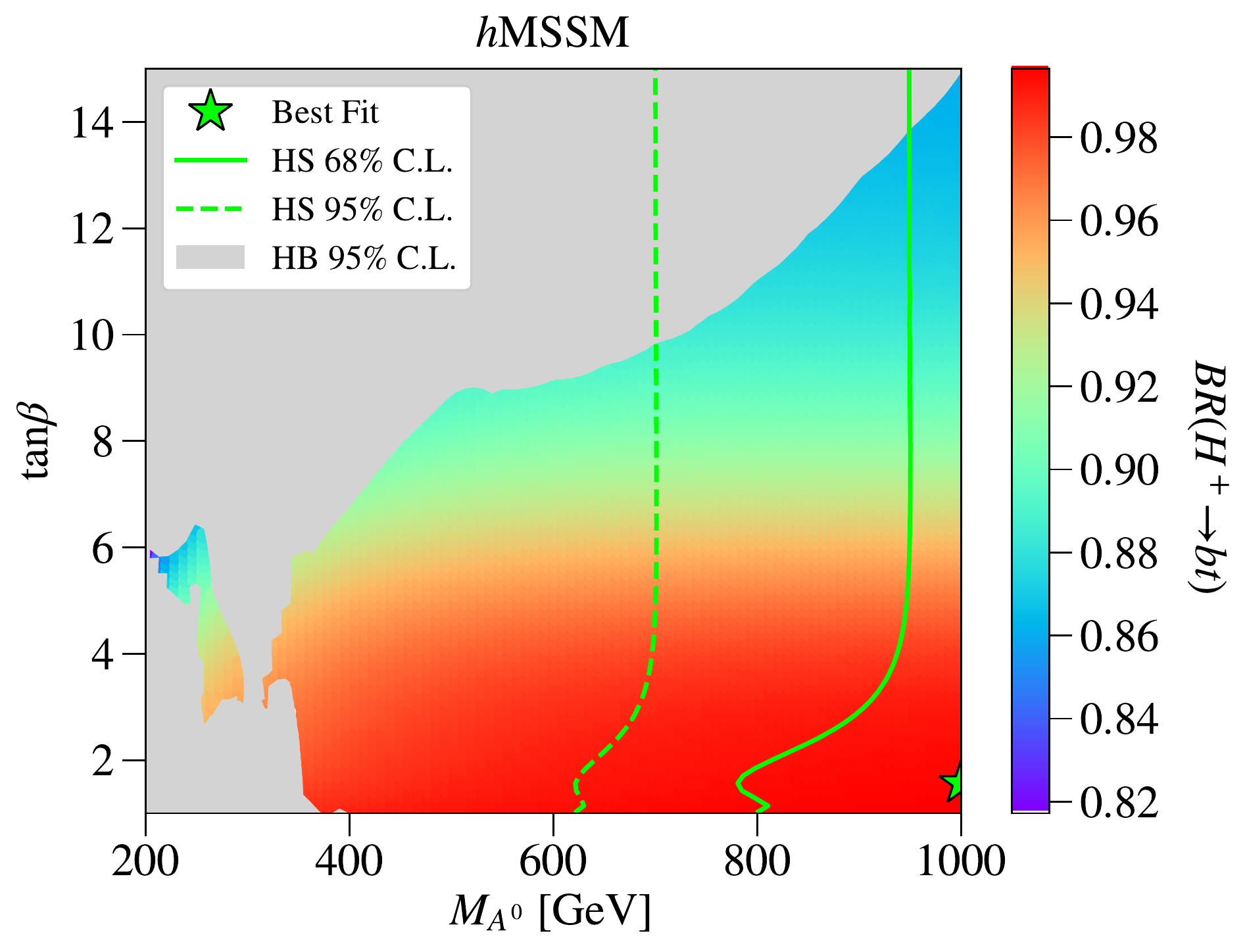}
\includegraphics[scale=0.45]{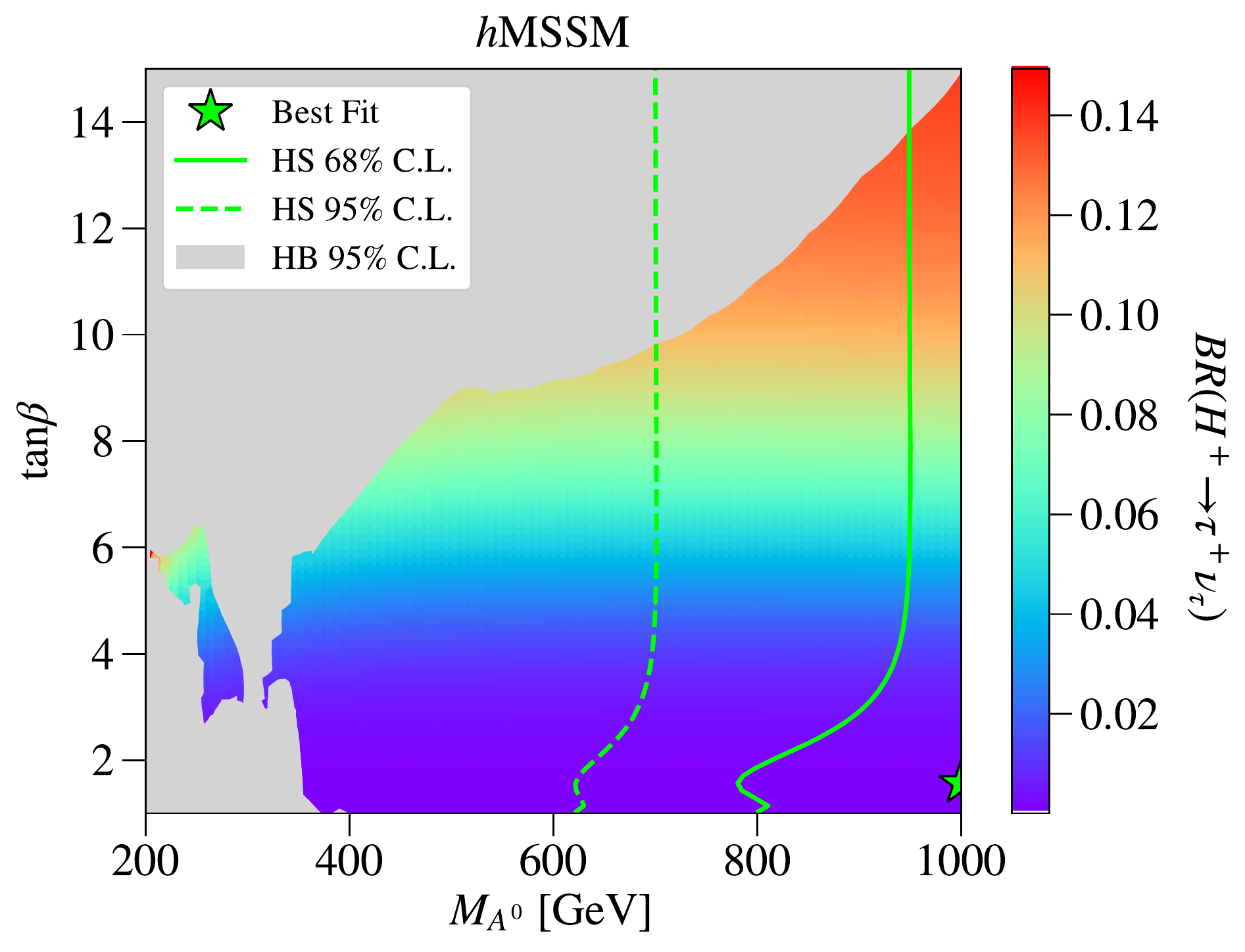}
\caption{Allowed regions from Fig.~\ref{fig:hMSSM_HB5HS2_excl} in the  ($M_{A^0}$, $\tan\beta$) plane 
for \hmssm scenario, with the color representing the BRs {\rm BR}($H^+ \to tb$) (left) and {\rm BR}($H^+ \to \tau \nu$) (right). }
\label{fig:hMSSM_HB5HS2_br}
\end{figure}

In the \hmssm scenario, the charged Higgs decays mainly into top-bottom with more than 90\% BR for 
$\tan\beta\leq 8$, see Fig.~\ref{fig:hMSSM_HB5HS2_br} (left), which decreases for larger $\tan\beta$ values. For small $\tan \beta$, the BR$(H^+ \to t \bar b)$ is very close
to 100\%. In this scenario, the $\tau\nu$ channel has a rather small
BR, less than 10\%, in most of the cases as depicted in Fig.~\ref{fig:hMSSM_HB5HS2_br} (right) and
becomes negligible for low $\tan \beta$.

\subsection{The \mhmodp case}
\begin{figure}[H]
\includegraphics[scale=0.45]{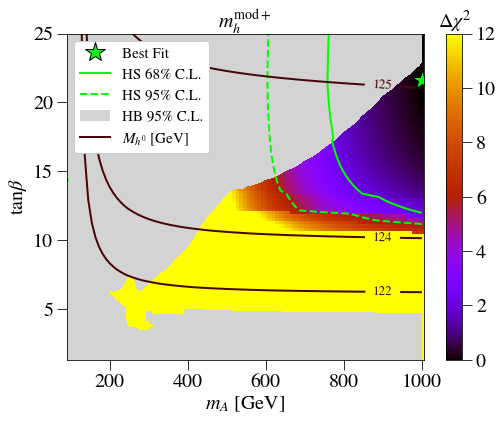}
\includegraphics[scale=0.45]{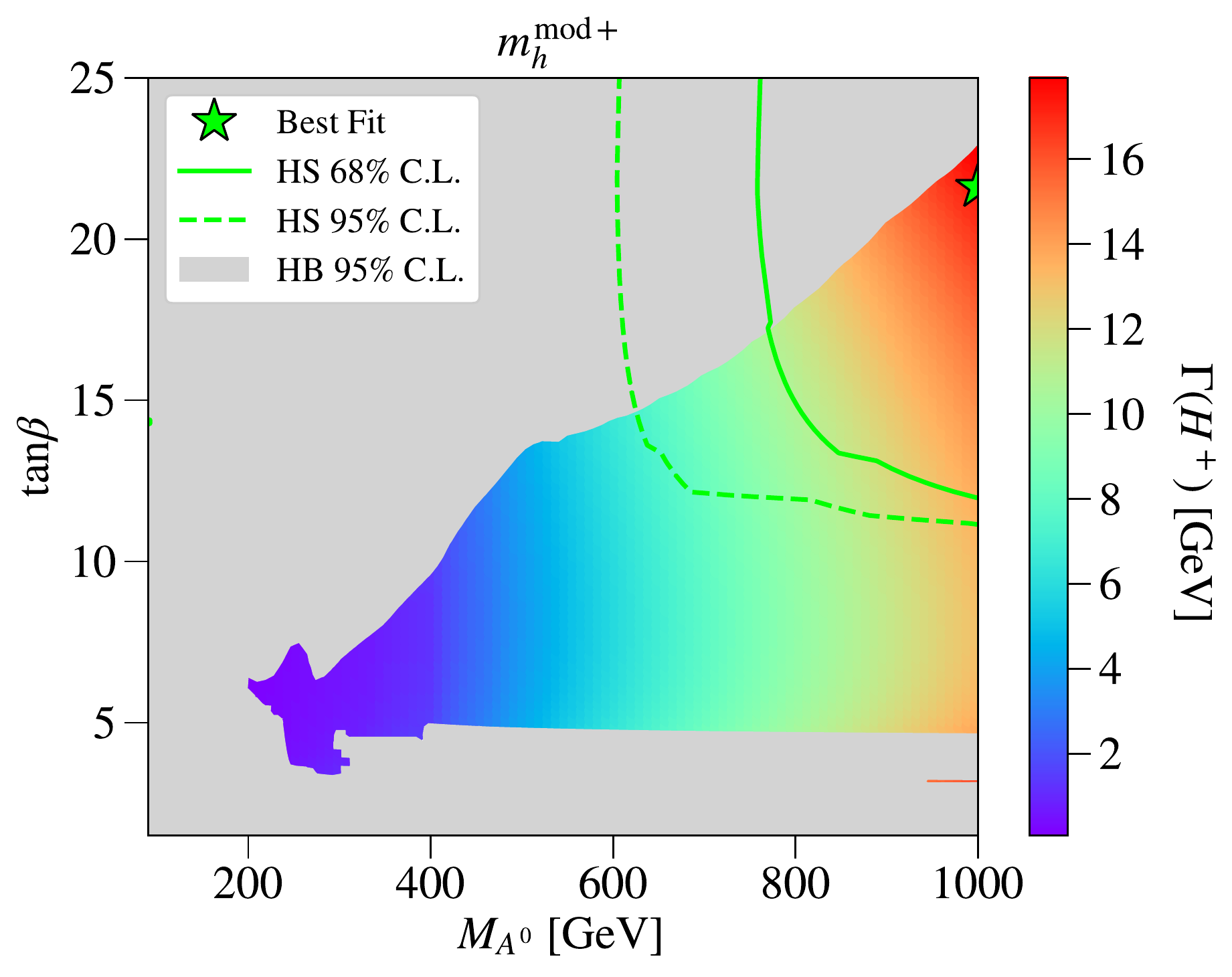}
\includegraphics[scale=0.45]{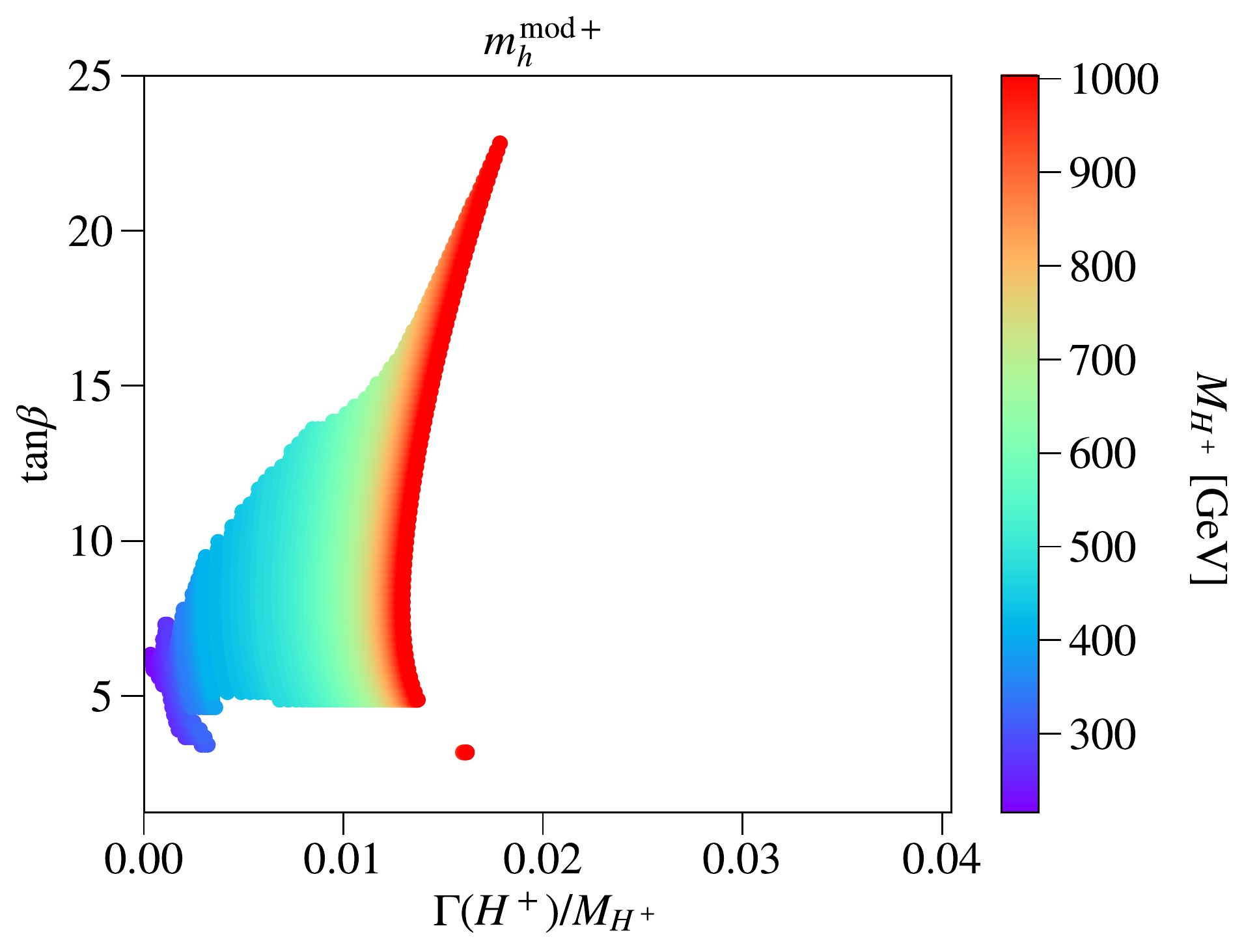}
\includegraphics[scale=0.45]{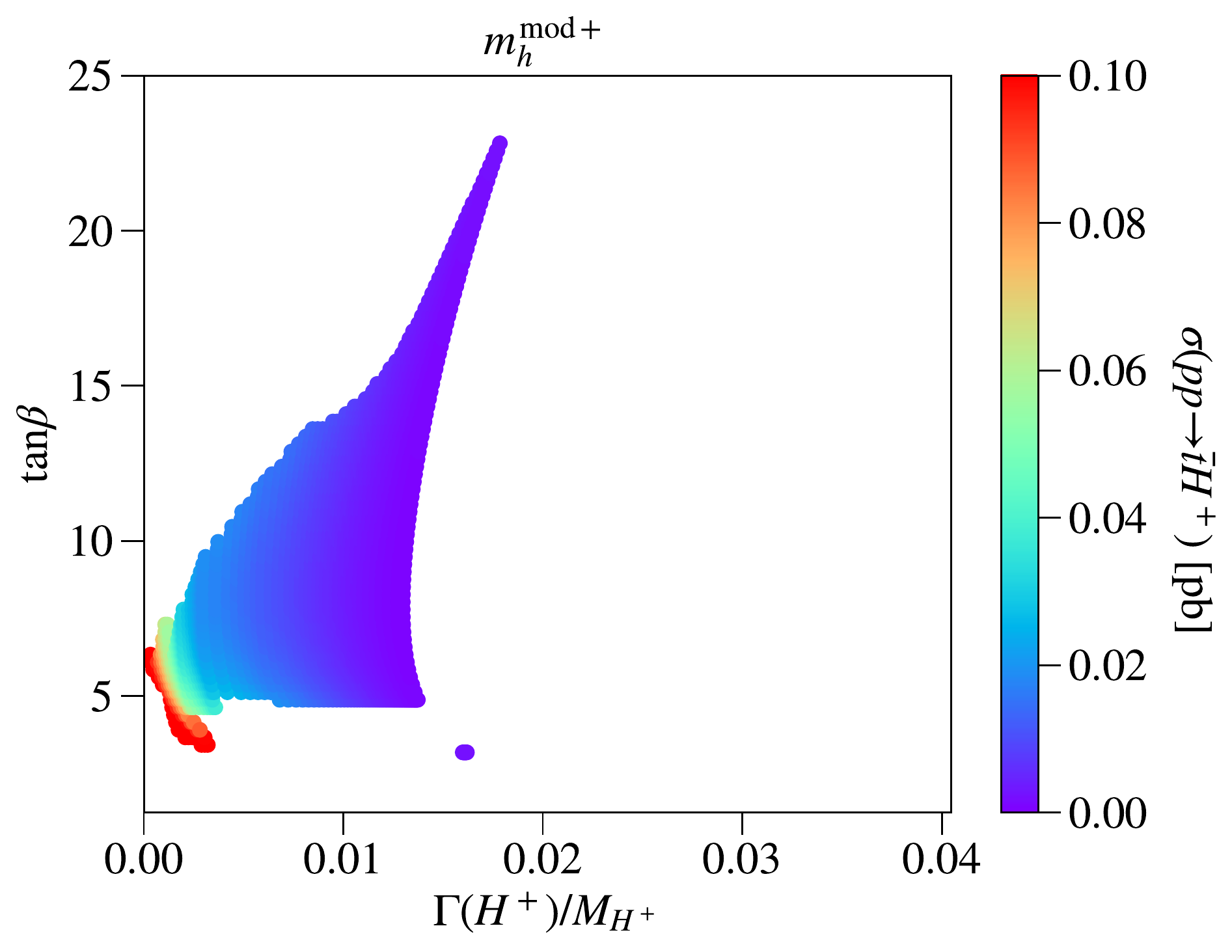}
\caption{
Allowed parameter region in the \mhmodp scenario over the ($m_A\equiv M_{A^0}$, $\tan\beta$) plane with colour showing $\Delta\chi^2$ (top-left)
and the charged Higgs boson mass (top-right). The LHC Higgs searches constraints are included. The light green contours are HiggsSignals exclusion 
limits at $1\sigma$ (solid) and $2\sigma$ (dashed). The light gray area is excluded by HiggsBounds at $2\sigma$.
The solid brown lines are contours for the lighter CP-even scalar $h^0$ mass. The best fit point is located at $M_{H^\pm}\approx 1$ TeV and $\tan\beta =20$.
In the two bottom panels of Fig. \ref{fig:mhmodp_HB5HS2_excl} we present $\tan \beta$ as a function
of  $\Gamma_{H^\pm}/M_{H^\pm}$ with the colour code showing the charged Higgs mass (left) and
the charged Higgs production cross section (right).  }
\label{fig:mhmodp_HB5HS2_excl}
\end{figure}

In the \mhmodp scenario, the allowed parameter region is shown in Fig. \ref{fig:mhmodp_HB5HS2_excl} (top-left).
The best fit point is located at $M_{H^\pm}\approx 1$ TeV and $\tan\beta =20$. In order
to have a low $\Delta\chi^2$ and simultaneously a light CP-even Higgs, close to 125 GeV, 
a value of $\tan\beta>10$ is required.  The latter requirement leads to a suppression of the total width of the charged Higgs because the BR to top-bottom
is proportional to $m_t/\tan\beta$. The same argument holds for the charged Higgs production cross section which
becomes smaller than in the previous scenario. The total charged Higgs width is shown in the allowed parameter
region in the top-right panel of Fig. \ref{fig:mhmodp_HB5HS2_excl}. Again, we need a heavy charged Higgs boson to obtain 
sizeable total widths but, contrary to the previous scenario, we now need quite large values of $\tan \beta$.
In the two bottom panels of Fig.~\ref{fig:mhmodp_HB5HS2_excl}, we present $\tan \beta$ as a function
of  $\Gamma_{H^\pm}/M_{H^\pm}$ with the colour showing the charged Higgs mass (left) and
the charged Higgs production cross section (right). Clearly, a compromise has to be reached between 
the values chosen for the charged Higgs mass while having non-negligible values for the production cross section. 

\begin{figure}[H]
        \includegraphics[scale=0.45]{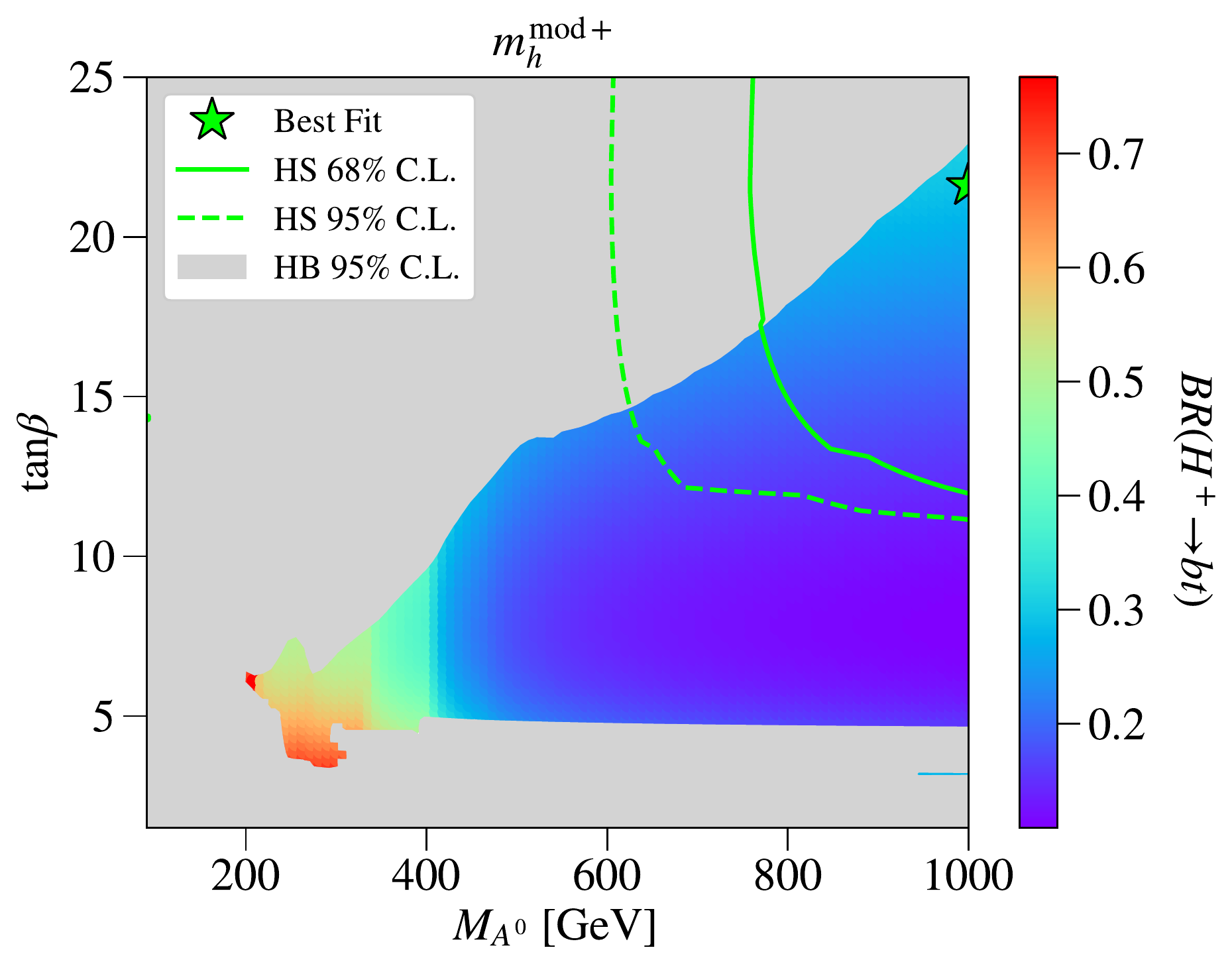}
        \includegraphics[scale=0.45]{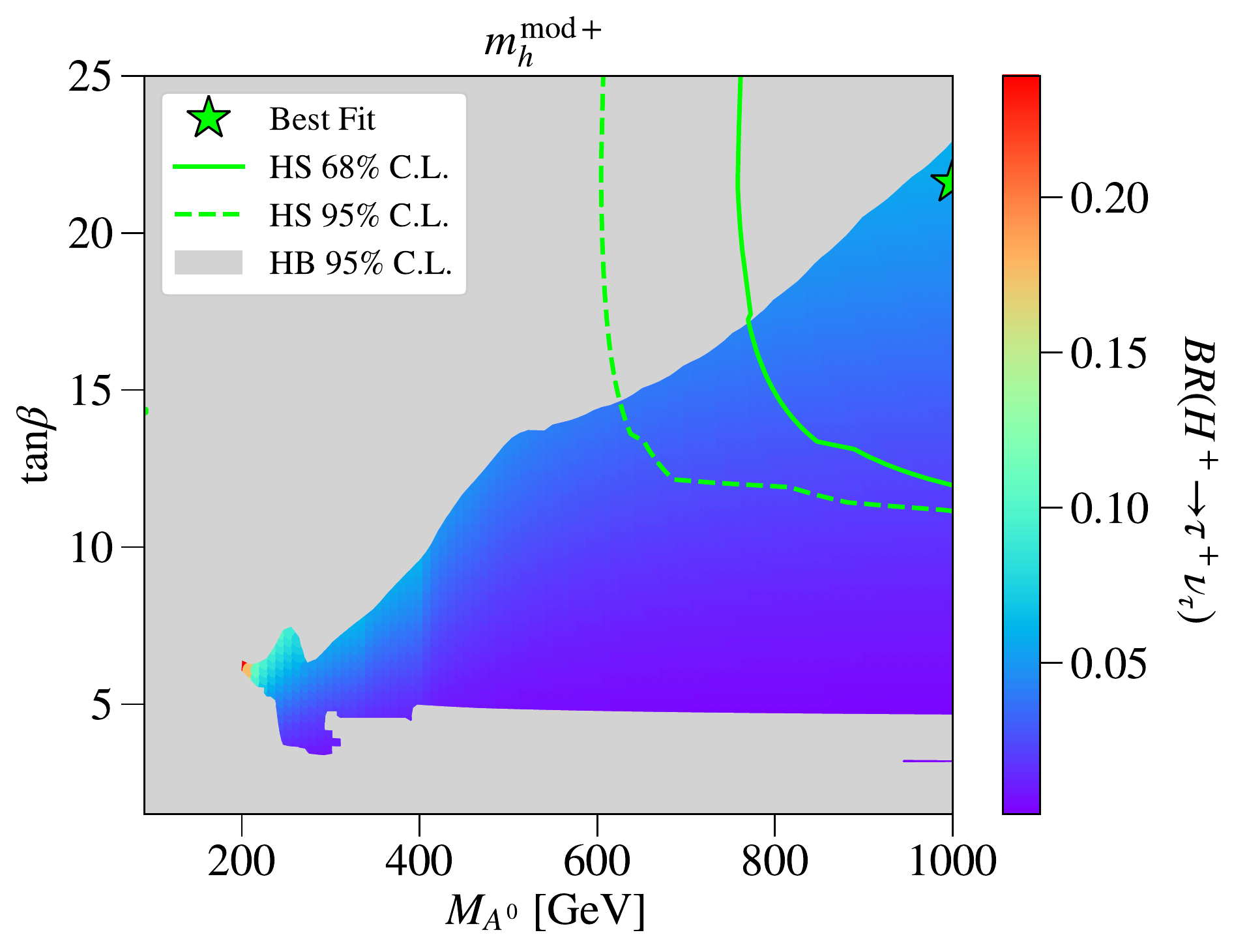}
        \includegraphics[scale=0.45]{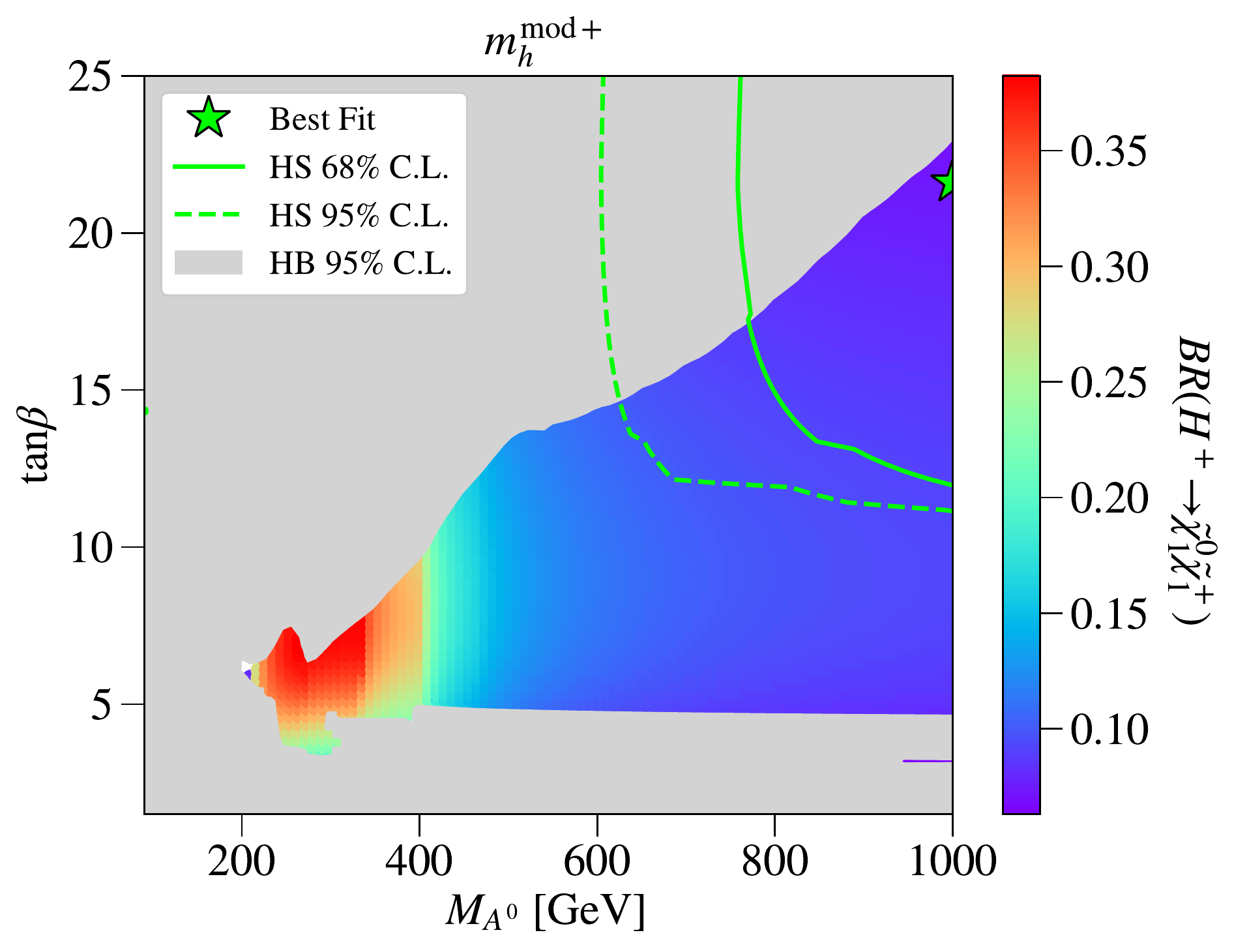}
        \includegraphics[scale=0.45]{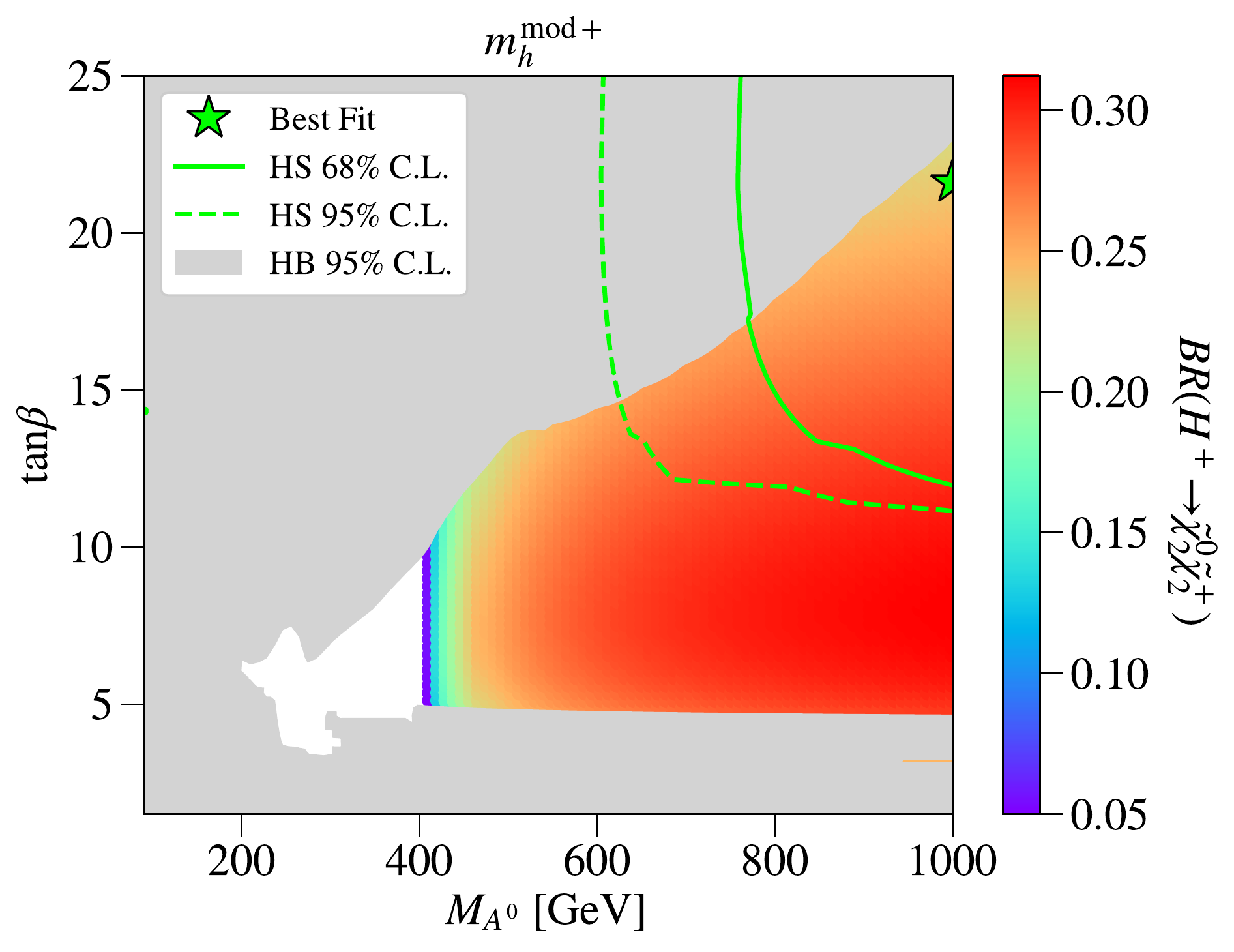}
\caption{Allowed regions, as shown in Fig. \ref{fig:mhmodp_HB5HS2_excl}, in the ($M_{A^0}$, $\tan\beta$) plane. 
We present the BR $H^+\to t\bar b$ (top-left),  $H^+\to \bar \tau \nu$ (top-right), 
 $H^+\to \chi^0_1\chi^+_1$ (bottom-left) and  $H^+\to \chi^0_2\chi^+_2$ (bottom-right).}
\label{fig:mhmodp_HB5HS2_BR}
\end{figure}

In Fig.~\ref{fig:mhmodp_HB5HS2_BR}, we again show the allowed region and the colour illustrates the charged Higgs BRs. 
In the top panels, we show  ${\rm BR}(H^+\to t\bar b)$ and  ${\rm BR}(H^+\to \bar \tau \nu)$ where it can be seen that 
the $tb$ BR is larger than the $\tau\nu$ one. In the bottom panels, we only illustrate the dominant chargino-neutralino channels,
namely$:$ ${\rm BR}(H^+\to \chi^0_1\chi^+_1)$  and  ${\rm BR}(H^+\to \chi^0_2\chi^+_2)$.

\subsection{Benchmark Points (BPs)}

This section briefly outlines the BPs found via the methodology outlined in the previous section. 

{In Tab.~\ref{tab:benchmarks_points_hmssm} we present four BPs for the $h$MSSM scenario with a value of $\tan\beta = 1.01$ and $5$ and with a range of charged Higgs masses between 275 and 633.91 GeV. BP4 in this table was chosen for the numerical analysis in the following section as it has a high charged Higgs width-to-mass ratio and thus is expected to have a high interference entering the signal cross section.}

{In Tab.~\ref{tab:benchmarks_points_mhmodp} we present four BPs for the $m_{h}^{\mathrm{mod}+}$ scenario for a number of values of $\tan\beta$ between 3.42 and 20 and charged Higgs masses between 303.08 and 900 GeV. BP4 was chosen for the numerical analysis of the next section because it has a high cross section even though the charged Higgs width-to-mass ratio is very low. This BP provides insight into the behaviour of interference throughout the cutflow in the scenario where interference is small relative to the signal.}

{In both scenarios, BPs 1 to 3 are presented to motivate further research of interference in the future as they should provide interesting and, as yet un-excluded, points of the MSSM.}

\begin{table}[!t]
    \centering
    {\renewcommand{\arraystretch}{1}
    {\setlength{\tabcolsep}{0.3cm}
    \begin{tabular}{|c|c|c|c|c|}
        \hline
        Parameters & BP1 & BP2 & BP3 & BP4 \\
        \hline
        \hline
        \multicolumn{5}{|c|}{MSSM inputs} \\
        \hline
        $\tan\beta$ & $5$ & $5$ & $1.01$ & $1.01$ \\
        \hline
        \multicolumn{5}{|c|}{Masses in GeV} \\
        \hline
        $M_{h^0}$ & $125$ & $125$ & $125$ & $125$ \\
        $M_{H^0}$ & $266.77$ & $495.19$ & $615.32$ & $648.3$ \\
        $M_{A^0}$ & $263$ & $493.5$ & $594.6$ & $628.79$ \\
        $M_{H^+}$ & $275.01$ & $500$ & $600.01$ & $633.91$ \\
        \hline
        \hline
        \multicolumn{5}{|c|}{Total decay width  in GeV} \\
        \hline
        $\Gamma(H^+)$ & $0.3499$ & $1.0423$ & $26.177$ & $27.777$ \\
        \hline
        \hline
        \multicolumn{5}{|c|}{BR$(H^+ \to XY)$ in \%} \\
        \hline
        BR$(H^+ \to b t)$ & $91.665$ & $96.105$ & $99.375$ & $99.418$ \\
        \hline
        \hline
        \multicolumn{5}{|c|}{Ratios} \\
        \hline
        $\Gamma(H^+)/M_{H^+}$ & $0.0012723$ & $0.0020846$ & $0.043628$ & $0.043819$ \\
        \hline
        \hline
        \multicolumn{5}{|c|}{Cross sections in pb} \\
        \hline
        $\sigma(pp \to \bar{t}H^+)$ & $0.0932$ & $0.0177$ & $0.2090$ & $0.1431$ \\
        $\sigma(pp \to \bar{t}H^+) \times BR(H^+ \to b t)$ & $0.0854$ & $0.0170$ & $0.2077$ & $0.1423$ \\
        \hline 
    \end{tabular}}}
    \caption{BPs for the $h\mathrm{MSSM}$ scenario.}
    \label{tab:benchmarks_points_hmssm}
\end{table}

\begin{table}[!t]
    \centering
    {\renewcommand{\arraystretch}{0.8}
    {\setlength{\tabcolsep}{0.3cm}
    \begin{tabular}{|c|c|c|c|c|}
        \hline
        Parameters & BP1 & BP2 & BP3 & BP4 \\
        \hline
        \hline
        \multicolumn{5}{|c|}{MSSM inputs} \\
        \hline
        $\tan\beta$ & $6$ & $10$ & $20$ & $3.42$ \\
        \hline
        \hline
        \multicolumn{5}{|c|}{Masses in GeV} \\
        \hline
        $M_{h^0}$ & $120.18$ & $122.46$ & $123.47$ & $113.55$ \\
        $M_{H^0}$ & $595.33$ & $695.4$ & $896.01$ & $298.69$ \\
        $M_{A^0}$ & $594.38$ & $695.12$ & $895.96$ & $292.22$ \\
        $M_{H^+}$ & $600$ & $700$ & $900$ & $303.08$ \\
        $M_{\tilde{\chi}^+_1}$ & $139.97$ & $144.16$ & $147.54$ & $133.2$ \\
        $M_{\tilde{\chi}^+_2}$ & $270.8$ & $268.59$ & $266.75$ & $274.19$ \\
        $M_{\tilde{\chi}^0_1}$ & $84.345$ & $86.404$ & $87.934$ & $80.637$ \\
        $M_{\tilde{\chi}^0_2}$ & $147.2$ & $149.46$ & $151.39$  & $143.88$ \\
        $M_{\tilde{\chi}^0_3}$ & $209.7$ & $209.8$ & $210.14$ & $205.33$ \\
        $M_{\tilde{\chi}^0_4}$ & $271.76$ & $268.81$ & $266.41$ & $276.42$ \\
        $M_{\tilde{b}_1}$ & $1000$ & $999.82$ & $996.08$ & $998.97$ \\
        $M_{\tilde{b}_2}$ & $1002$ & $1002.2$ & $1006$ & $1002.8$ \\
        $M_{\tilde{t}_1}$ & $876.49$ & $876.45$ & $876.43$ & $876.61$ \\
        $M_{\tilde{t}_2}$ & $1134.8$ & $1134.8$ & $1134.8$ & $1134.9$ \\
        \hline
        \hline
        \multicolumn{5}{|c|}{Total decay width  in GeV} \\
        \hline
        $\Gamma(H^+)$ & $5.8582$ & $7.7229$ & $14.311$ & $0.9253$ \\
        \hline
        \hline
        \multicolumn{5}{|c|}{BR$(H^+ \to XY)$ in \%} \\
        \hline
        BR$(H^+ \to \tilde{\chi}^0_1 \tilde{\chi}^+_1)$ & $10.789$ & $10.379$ & $7.7896$ & $20.73$ \\
        BR$(H^+ \to \tilde{\chi}^0_2 \tilde{\chi}^+_2)$ & $27.858$ & $29.296$ & $24.307$ & $-$ \\
        BR$(H^+ \to \tilde{\chi}^+_1 \tilde{\chi}^0_3)$ & $13.003$ & $12.161$ & $9.1983$ & $-$ \\
        BR$(H^+ \to \tilde{\chi}^+_1 \tilde{\chi}^0_4)$ & $18.454$ & $18.648$ & $15.061$ & $-$ \\
        BR$(H^+ \to \tilde{\chi}^0_3 \tilde{\chi}^+_2)$ & $9.7934$ & $11.002$ & $9.5996$ & $-$ \\
        BR$(H^+ \to \tau^+ \nu_{\tau})$ & $0.73738$ & $1.8127$ & $5.031$ & $0.7682$\\
        BR$(H^+ \to b t)$ & $15.728$ & $13.718$ & $26.989$ & $72.036$ \\
        \hline
        \hline
        \multicolumn{5}{|c|}{Ratios} \\
        \hline
        $\Gamma(H^+)/M_{H^+}$ & $0.0097637$ & $0.011033$ & $0.015901$ & $0.0031$ \\
        \hline
        \hline
        \multicolumn{5}{|c|}{Cross sections in pb} \\
        \hline
        $\sigma(pp \to \bar{t}H^+)$ &                                                      $0.007120$ & $0.003170$ & $0.002850$ & $0.130750$ \\
        $\sigma(pp \to \bar{t}H^+) \times BR(H^+ \to \tilde{\chi}^0_1 \tilde{\chi}^+_1)$ & $0.000768$ & $0.000329$ & $0.000222$ & $0.027100$ \\
        $\sigma(pp \to \bar{t}H^+) \times BR(H^+ \to \tilde{\chi}^0_2 \tilde{\chi}^+_2)$ & $0.001984$ & $0.000929$ & $0.000693$ & $-$ \\
        $\sigma(pp \to \bar{t}H^+) \times BR(H^+ \to \tilde{\chi}^+_1 \tilde{\chi}^0_3)$ & $0.000926$ & $0.000386$ & $0.000262$ & $-$ \\
        $\sigma(pp \to \bar{t}H^+) \times BR(H^+ \to \tilde{\chi}^+_1 \tilde{\chi}^0_4)$ & $0.001314$ & $0.000591$ & $0.000429$ & $-$ \\
        $\sigma(pp \to \bar{t}H^+) \times BR(H^+ \to \tilde{\chi}^0_3 \tilde{\chi}^+_2)$ & $0.000697$ & $0.000349$ & $0.000274$ & $-$ \\
        $\sigma(pp \to \bar{t}H^+) \times BR(H^+ \to \tau^+ \nu_{\tau})$ &                 $0.000053$ & $0.000057$ & $0.000143$ & $-$ \\
        $\sigma(pp \to \bar{t}H^+) \times BR(H^+ \to b t)$ &                               $0.001120$ & $0.000435$ & $0.000769$ & $0.094200$ \\
        \hline
    \end{tabular}}}
    \caption{BPs for $m_{h}^{\mathrm{mod}+}$ scenario.}
    \label{tab:benchmarks_points_mhmodp}
\end{table}

\section{Results}
\begin{figure}[!t]
\centering
\includegraphics[scale=0.8]{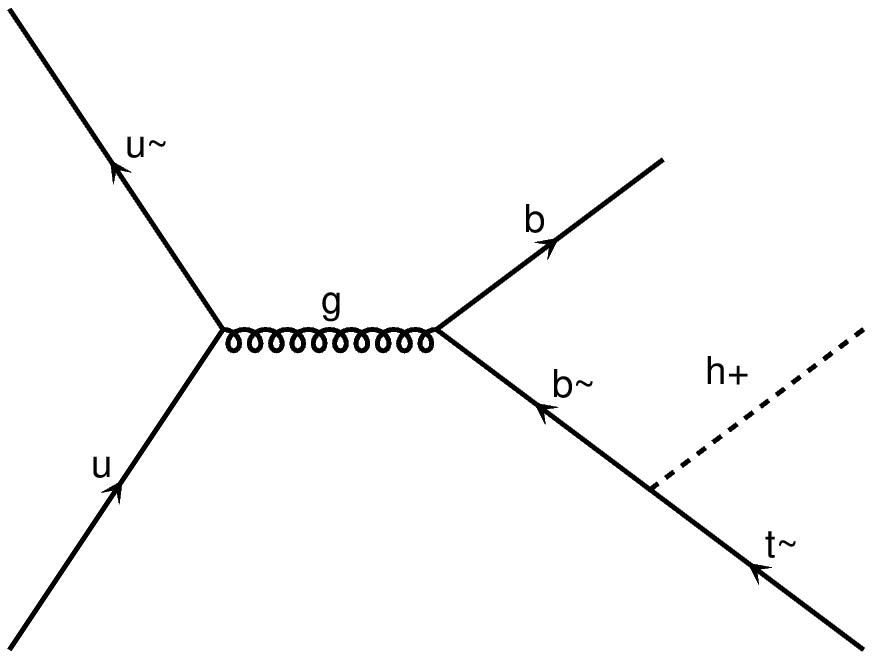}
\includegraphics[scale=0.8]{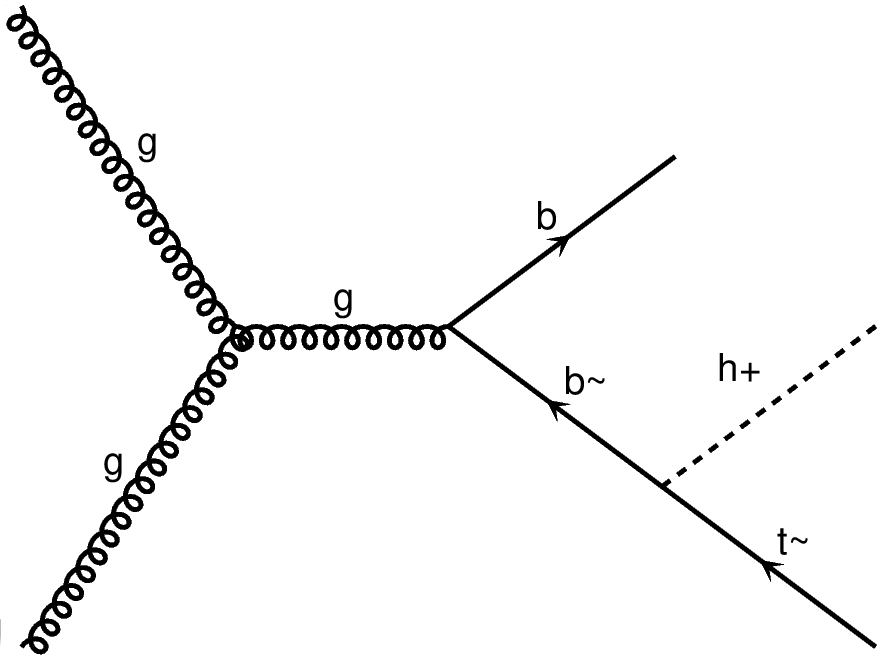} \\
\includegraphics[scale=0.8]{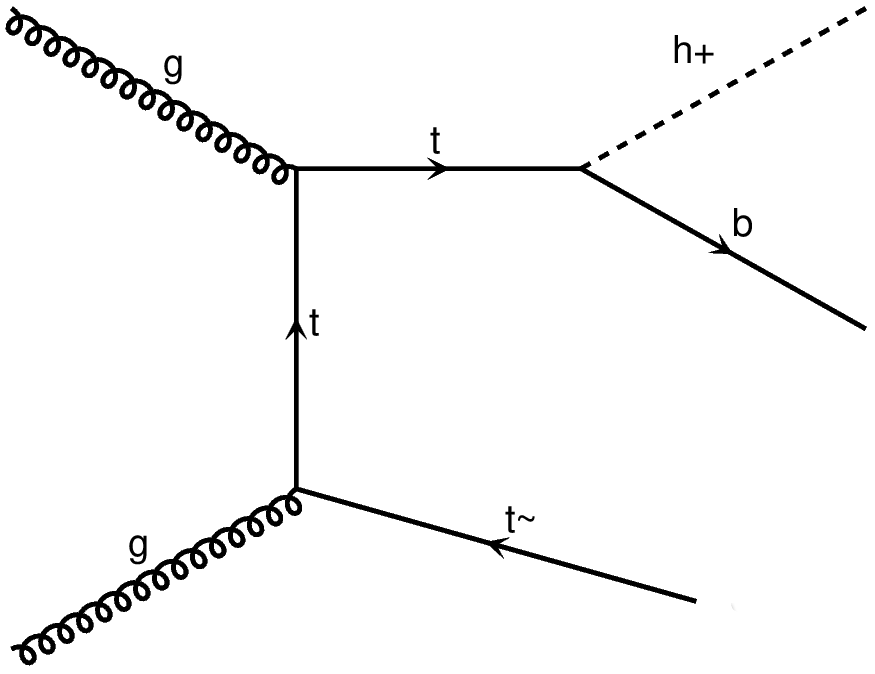}
\includegraphics[scale=0.175]{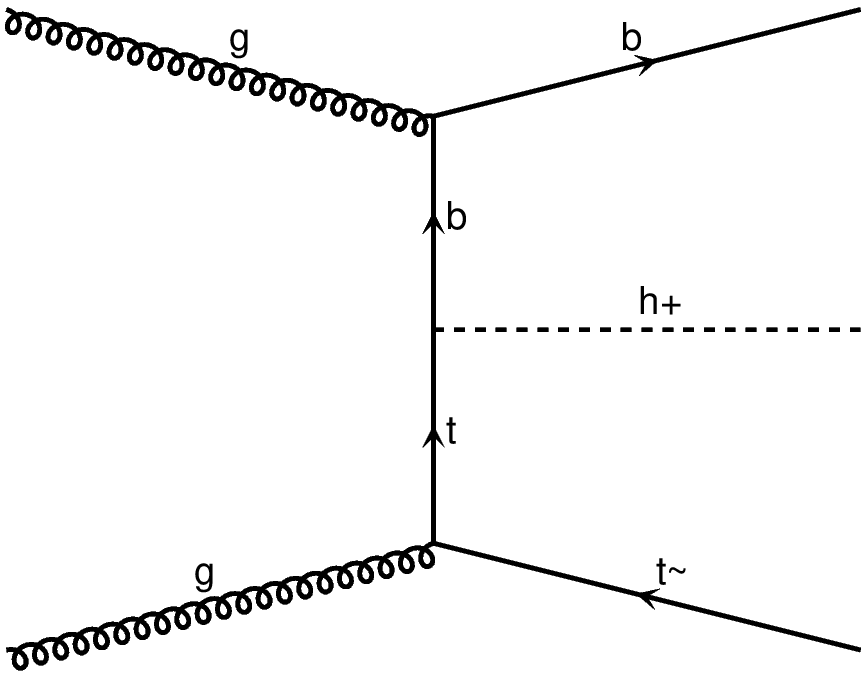}
\caption{\label{fig:signal-diagrams} A selection of signal Feynman diagrams.}
\end{figure}

\begin{figure}[!ht]
\centering
\includegraphics[scale=0.7]{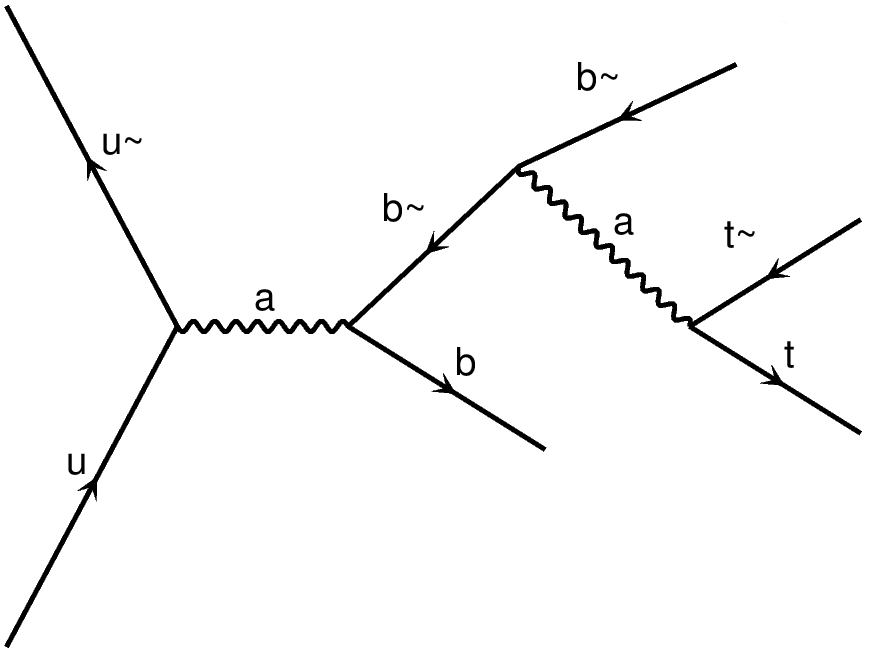}
\includegraphics[scale=0.7]{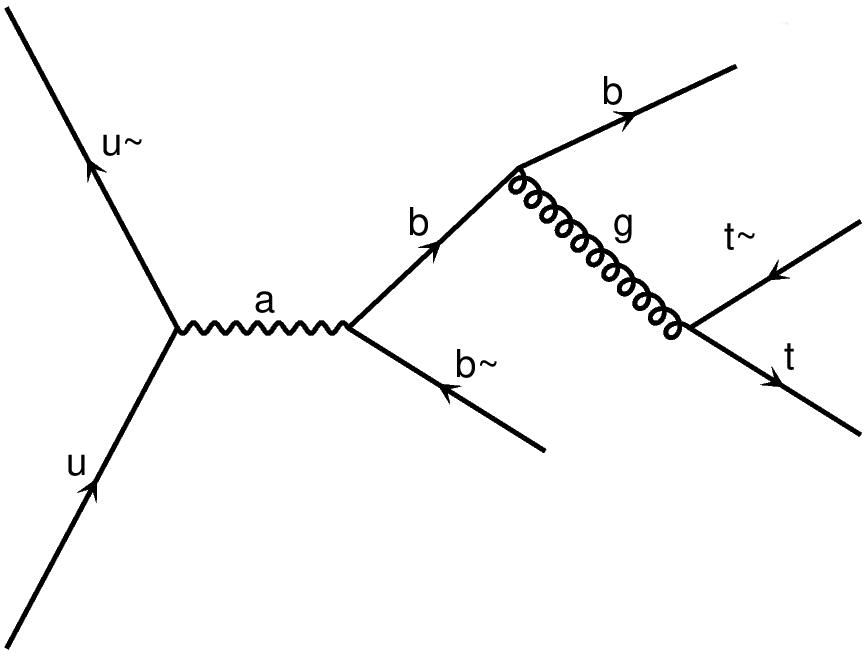}
\includegraphics[scale=0.7]{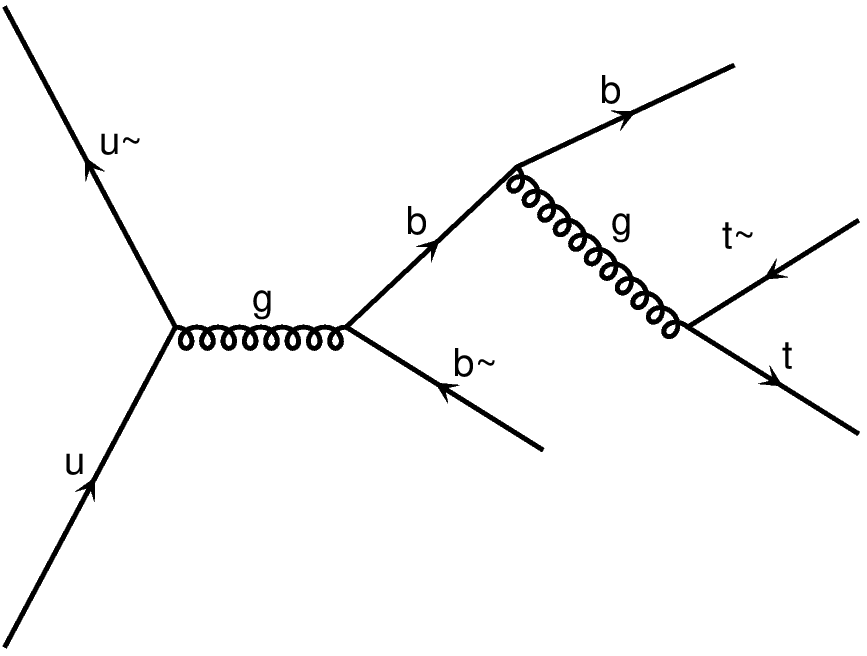} \\

\includegraphics[scale=0.225]{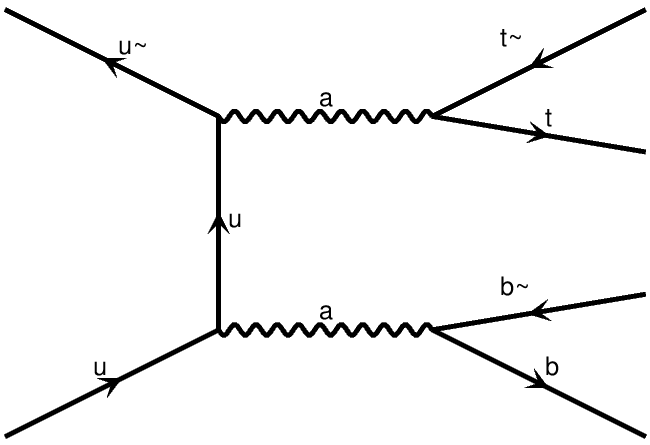} \hspace{0cm}
\includegraphics[scale=0.225]{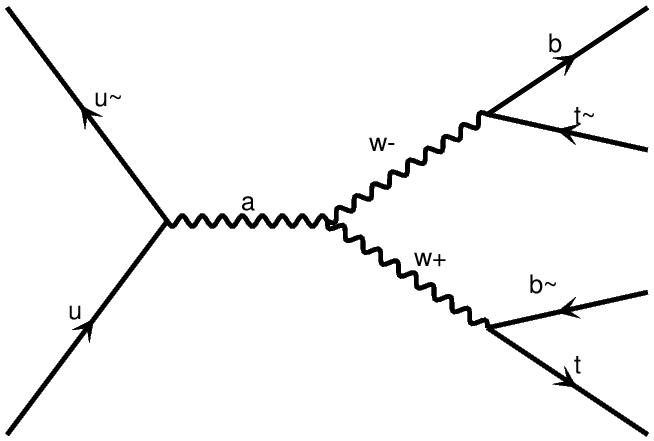} \hspace{0cm}
\includegraphics[scale=0.95]{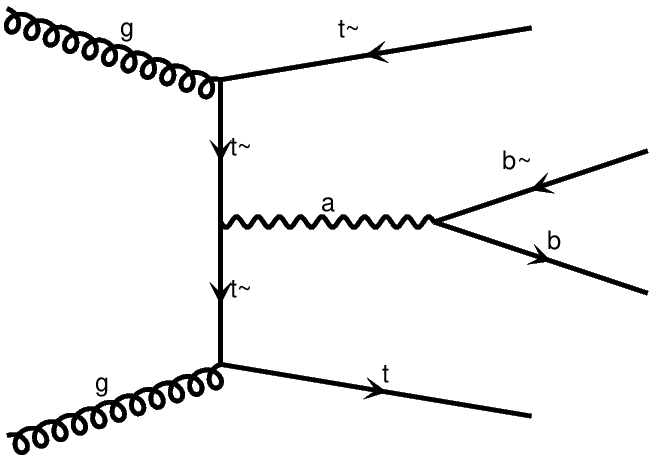} \hspace{0cm} \\ 
\includegraphics[scale=0.225]{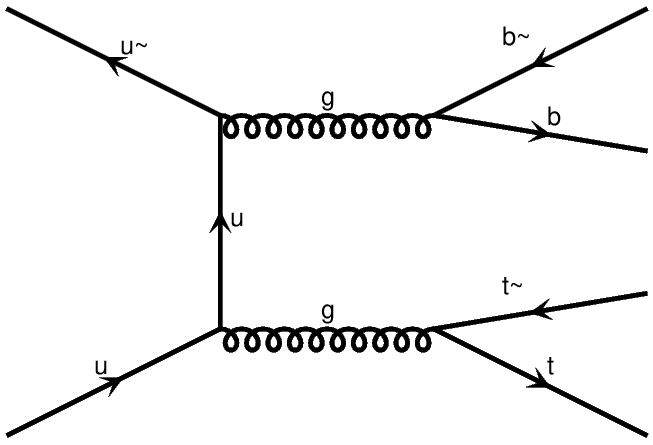}
\includegraphics[scale=0.225]{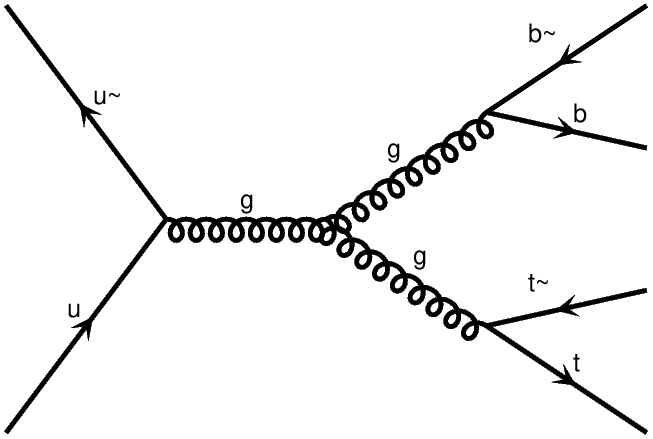}
\includegraphics[scale=0.7]{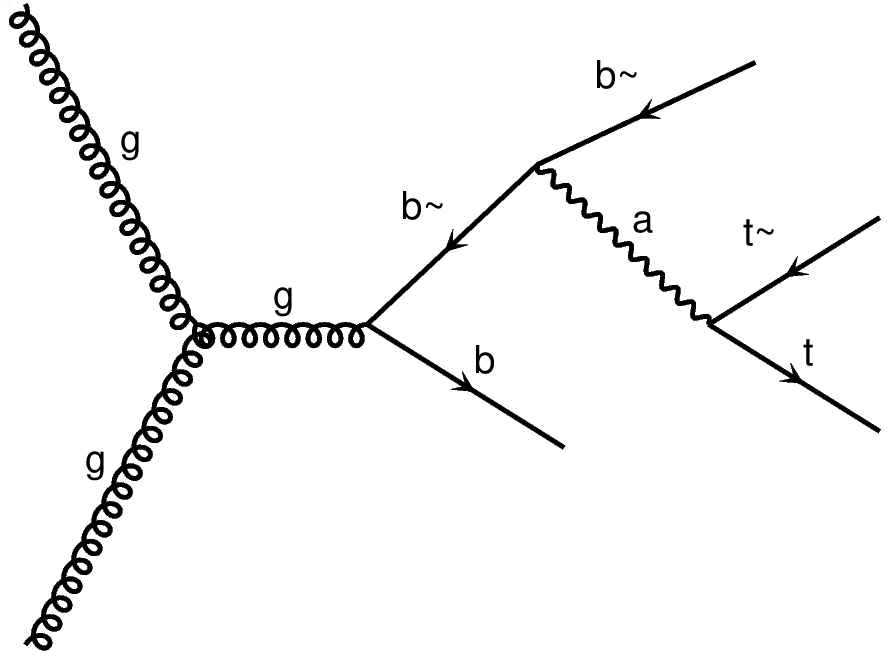} 

\caption{\label{fig:background-diagrams} A selection of background Feynman diagrams.}
\end{figure}

{As intimated, the process studied at MC level  is $p p \to t\bar{b}H^{-} \to tb\bar{t}\bar{b}$ (+ c.c.), thus the signal is defined as all processes in the MSSM mediated by the charged Higgs with a $t\bar{t}b\bar{b}$ final state while the background is defined as all processes in the MSSM with the same final state which are not mediated by a charged Higgs state. Figs.~\ref{fig:signal-diagrams} and \ref{fig:background-diagrams} present some examples of signal and background diagrams.}

Let us then define the scattering interference as $\rm I = \rm T-\rm S-\rm B$ where `$\rm T = \rm Total$' is the full scattering amplitude including all signal and background Feynman diagrams and the interference of these diagrams. `$\rm S$' is the signal scattering amplitude including only the signal diagrams  and `$\rm B$' is the background scattering amplitude including only the background diagrams. As the same phase space is shared by all of these terms, we can perform the calculation of these terms independently and evaluate the interference via the equation presented above.	

In order to explore the effects of interference on the search for a charged Higgs, we utilise BP4 found in Tab.~\ref{tab:benchmarks_points_hmssm} for the $h$MSSM case and BP4 found in Tab.~\ref{tab:benchmarks_points_mhmodp} for the $m^{\rm mod+}_{\rm h}$ case. These two points provide two kinematically distinct scenarios, one of which - the $h$MSSM one - has a high width-to-mass ratio for the charged Higgs boson, of $4.4$\%, while the other has a much lower ratio, of $0.31$\%.

Signal cross sections are significantly smaller than background cross sections before cuts. Hence the simulation of the $\rm T$ and $\rm B$ terms outlined above must have low uncertainty, which requires very large MC samples. This mandates a prudent use of computing resources and thus an extremely large sample of events was generated for $\rm T$, $\rm B$ and $\rm S$ at parton level to obtain a value for the cross section of these processes with very low MC error. After this was done, smaller detector level samples were generated for the purpose of applying cuts and obtaining efficiencies. The parton level cross sections and the detector level efficiencies are then used together to form the cutflow that will be presented in this section. The parton level results for both BPs can be found in Tab.~\ref{tab:partonresults}. {The error presented here is the MC error and  is displayed to show that a large enough parton level sample was generated to produce a sufficiently low uncertainty on the interference cross section. There are of course other sources of errors against which to tension the size of our interference effects, such as systematic theoretical errors due to the  finite order perturbative modelling of the signal and background cross sections, which can be large, especially for the latter (e.g.,   Ref.~\cite{Aaboud:2018cwk} finds that for the $t\bar t b\bar b$ background such an error is of order 10\%). Further, one ought to consider the systematic errors coming with the choice of Parton Distribution Functions (PDFs) and of their factorisation/renormalisation scale, which are expected to be similar in size. Hence, one will truly need to worry about the systematic error due to the presence of the interference effects studied here when they are beyond the 10\% or so level.}
 
\begin{table}[!ht]
		\centering
		\begin{tabular}{|lrrrrr|}
		\hline
	    {Model}  &  & {S} {(pb)} &  {B} {(pb)} &  {T} {(pb)} &  {I} {(pb)}  \\\hline
		{$h$MSSM}  & $\sigma$       & $0.03240$ & $13.078$ & $13.139$ & $0.028$ \\\hline
		 			    & $\Delta\sigma$ & $1.4\times 10^{-5}$ & $0.002$ & $0.001$ & $0.003$ \\\hline
		$m^{\rm{mod+}}_{{h}}$  & $\sigma$       & $0.08854$ & $13.095$ & $13.197$ & $0.014$ \\\hline
		                & $\Delta\sigma$ & $3.3\times 10^{-5}$ & $0.001$ & $0.001$ & $0.002$ \\\hline
		\end{tabular}
	\caption{\label{tab:partonresults} Parton level results for the $h$MSSM and $m^{\rm mod+}_{\rm h}$ benchmarks.}
\end{table}

The parton level sample for both scenarios contained 20,000,000 events generated in {MadGraph5}~\cite{Alwall:2014hca} at leading order with a Centre-of-Mass (CoM) energy of $13$ TeV, while the detector level samples for the hMSSM sample contained 5,000,000 events comprised of 100 independent samples of 50,000 events and the $m_{h}^{\rm mod+}$ sample contained 10,000,000 events comprised of 200 independent samples of 50,000 events. Both were generated in {MadGraph5} at leading order and at 13 TeV CoM energy. The detector level samples were then sent to {Pythia8}~\cite{Sjostrand:2014zea} for hadronisation/fragmentation and finally passed to {Delphes}~\cite{deFavereau:2013fsa} for detector smearing {utilising the standard ATLAS card}. Previous sections of this work calculated cross sections at NLO, however, as previously explained, this is not feasible for the background samples, so the MC analysis was undertaken at LO only. All samples included the decay of the charged Higgs, $H^{+} \to t\bar{b}$, to maximize statistics.

{Typical detector acceptances were utilised, namely electrons and muons must have transverse momentum $p_T > 7$ GeV and pseudo-rapidity $|\eta| < 2.5$ with 100\% lepton selection efficiency assumed. Jets must have $p_T > 20$ GeV and $|\eta| < 2.5$. Anti-k$_T$ jet clustering\cite{Cacciari:2008gp} was used and a $b$-tagging efficiency of 77\% and mis-tagging efficiency of 1\% employed.} We demand exactly one lepton in the final state, so that the longitudinal momentum of the missing energy can be solved for via 
\begin{equation}
p_\nu^z = \frac{1}{2p_{\ell T}^2}\left(A_Wp^z_{\ell}\pm E_\ell\sqrt{A_W^2\pm 4p^2_{\ell T}E^2_{\nu T}}\right),
\end{equation}
 where, $A_W = M^2_{W^{\pm}} + 2p_{\ell T}\cdot E_{\nu T}$. 

Reconstruction was then undertaken via the simultaneous minimisation of the following equations by permuting through all combinations of jets in the process,
				\begin{align}
				\chi^2_{\rm had} = \frac{\left(M_{\ell\nu} - M_W\right)^2}{\Gamma^2_W} 
									+\frac{\left(M_{jj} - M_W\right)^2}{\Gamma^2_W}		
									+\frac{\left(M_{\ell\nu j} - M_T\right)^2}{\Gamma^2_T}
								    +\frac{\left(M_{jjj} - M_T\right)^2}{\Gamma^2_T} 
									+\frac{\left(M_{jjjj} - M_{H^{\pm}}\right)^2}{\Gamma^2_{H^{\pm}}}
				\end{align} 
and
				\begin{align}
				\chi^2_{\rm lep} = \frac{\left(M_{\ell\nu} - M_W\right)^2}{\Gamma^2_W} 
									+\frac{\left(M_{jj} - M_W\right)^2}{\Gamma^2_W}		
									+\frac{\left(M_{\ell\nu j} - M_T\right)^2}{\Gamma^2_T}
								    +\frac{\left(M_{jjj} - M_T\right)^2}{\Gamma^2_T} 
									+\frac{\left(M_{\ell\nu jj} - M_{H^{\pm}}\right)^2}{\Gamma^2_{H^{\pm}}}
				\end{align}
	
The results of this reconstruction can be found in Figs.~\ref{fig:hmssm-recon-variables} and \ref{fig:mhmod-recon-variables}, normalised to unit area. {This reconstruction requires one to use the width of the particles, which introduces a model dependence. Thus, in the aforementioned figures, we also present the same reconstruction methodology but without the use of particle widths to highlight how this affects the reconstruction. We refer to these methodologies as the model dependent and model independent reconstructions, respectively.}

We apply a simple set of acceptance cuts to illustrate the sensitivity to the interference term, these cuts include a final state definition of 1 lepton, 5 or more jets, more than 2 or 3 $b$-jets, greater than 20 GeV missing transverse energy and, finally, the transverse mass of missing energy and the lepton must be higher than 60 GeV. Specifically, $m^{W}_{T} = \sqrt{ (\slashed{E}_x + \ell_x)^2 + (\slashed{E}_y + \ell_y)^2} > 60$ GeV. {It is an interesting question to experimentalists as to how interference contributions change with respect to varying $b$-tagging. This motivates the usage of 2 and 3 $b$-tags regions even if the increased $b$-tagging does not necessarily lead to increased signal significance.} This cutflow, applied to each of the BPs, can be found in Tabs.~\ref{tab:hmssmresults} and \ref{tab:mhmodresults}.

\subsection{The $h$MSSM analysis}
\begin{figure}[!ht]
\centering
\includegraphics[scale=0.3]{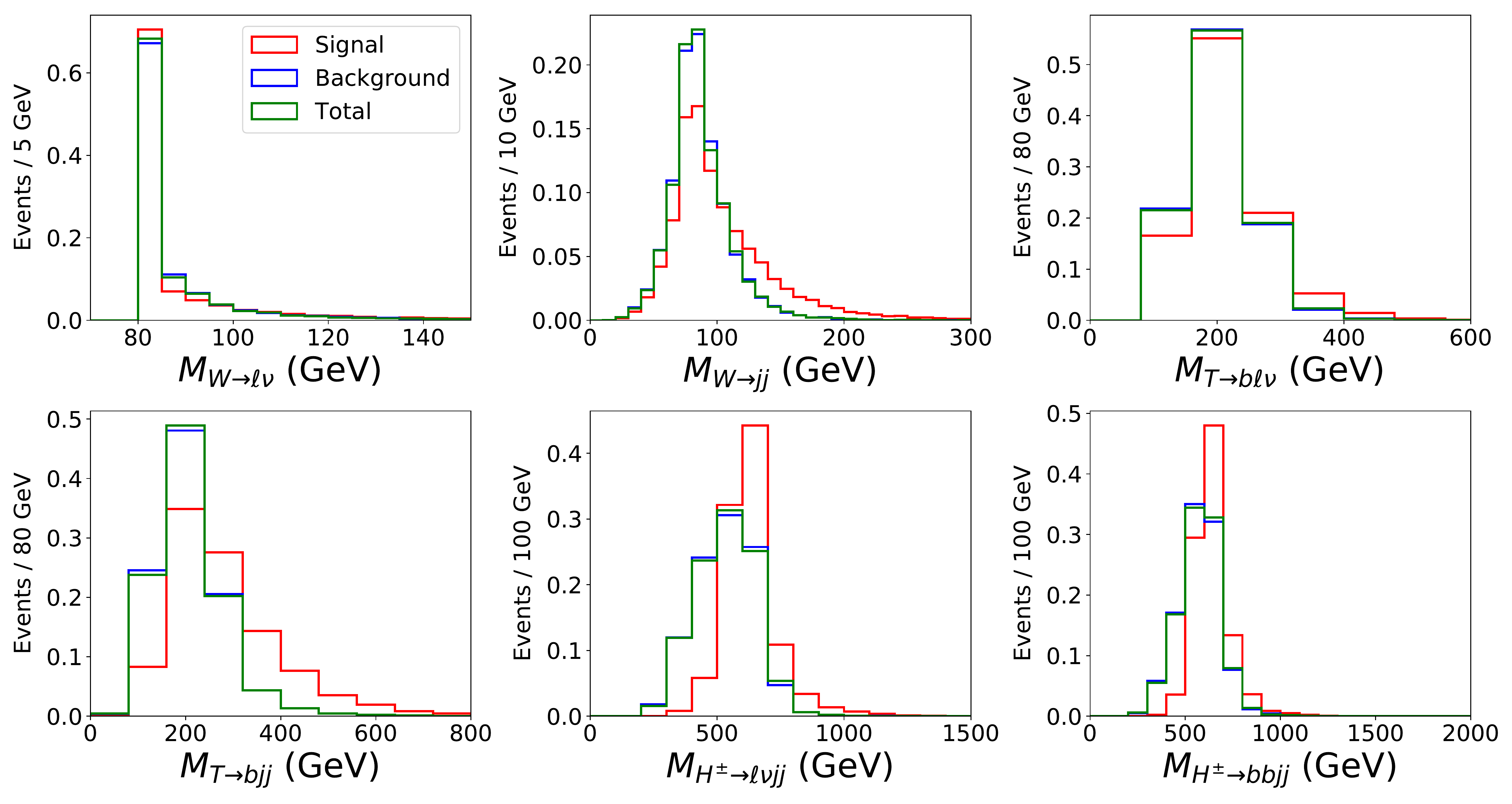}
\includegraphics[scale=0.3]{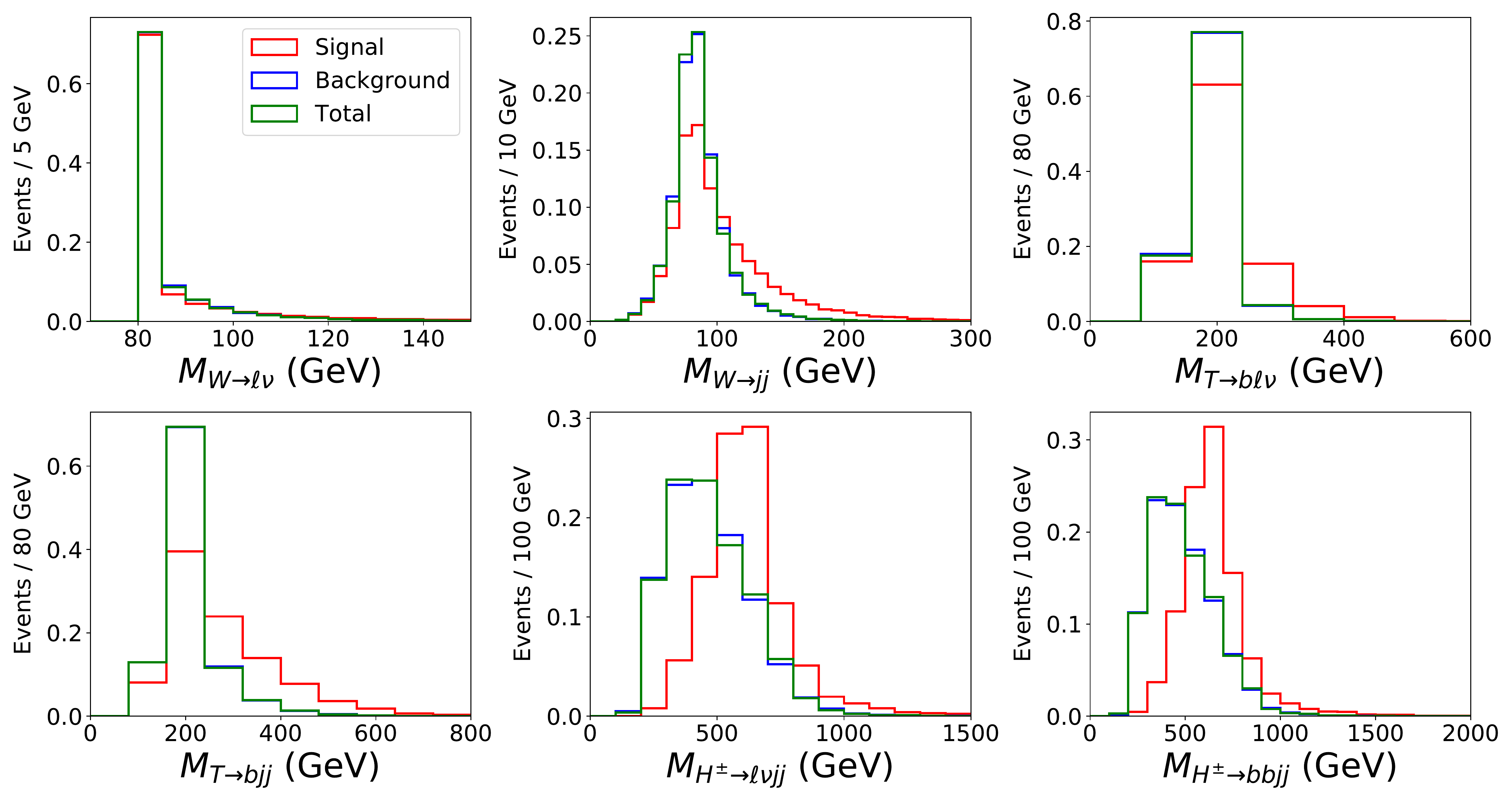}
\caption{\label{fig:hmssm-recon-variables} Invariant mass distributions for reconstructed particles in the $h$MSSM benchmark. Top: Model independent. Bottom: Model dependent.}
\end{figure}

It can be seen in Fig.~\ref{fig:hmssm-recon-variables} that all particles appear to be reconstructed very well. The model-dependent reconstruction and the model-independent case perform equally well for the signal. However, for the background and total samples the reconstruction is quite different. The model-dependent assumption provides a much better separation from the signal, this is especially apparent in both the leptonic and hadronic charged Higgs invariant mass distributions.

The ratio of signal cross section to interference cross section before cuts is $86.7$\%. This is an alarmingly high level of interference that a traditional experimental study would not account for correctly. The ratio after cuts in both the $\geq 2$ $b$-tag scenario and $\geq 3$ $b$-tag scenarios is $103.3$\% and $85.5$\% respectively, both extremely large interferences showing that the cutflow has done little to mitigate the magnitude of the interference relative to the signal. It should be noted that the uncertainty on the values in the $\geq 3$ $b$-tag region are approaching the same magnitude as the interference itself, thus strong conclusions in this region cannot be made.

It is important to note that the true effect of the interference is predicated on the overall shape of the interference distribution relative to the signal distribution. In general there are three cases~\cite{GAEMERS:1984347,DICUS:1994126,PhysRevD.93.055035}:
\begin{enumerate}
\item The interference takes the same shape as the signal and is positive, here we can expect a boosting of our new physics effects.
\item The interference takes the same shape as the signal and is negative, here we expect a cancellation of our new physics effects.
\item The interference takes a different shape and is either positive or negative, here we can expect a boosting and cancellation of new physics effects in different regions of phase space, manifesting as a ``peak-dip'' structure in the expected distributions.
\end{enumerate}

\begin{figure}[!ht]
	\centering
	\includegraphics[scale=0.3]{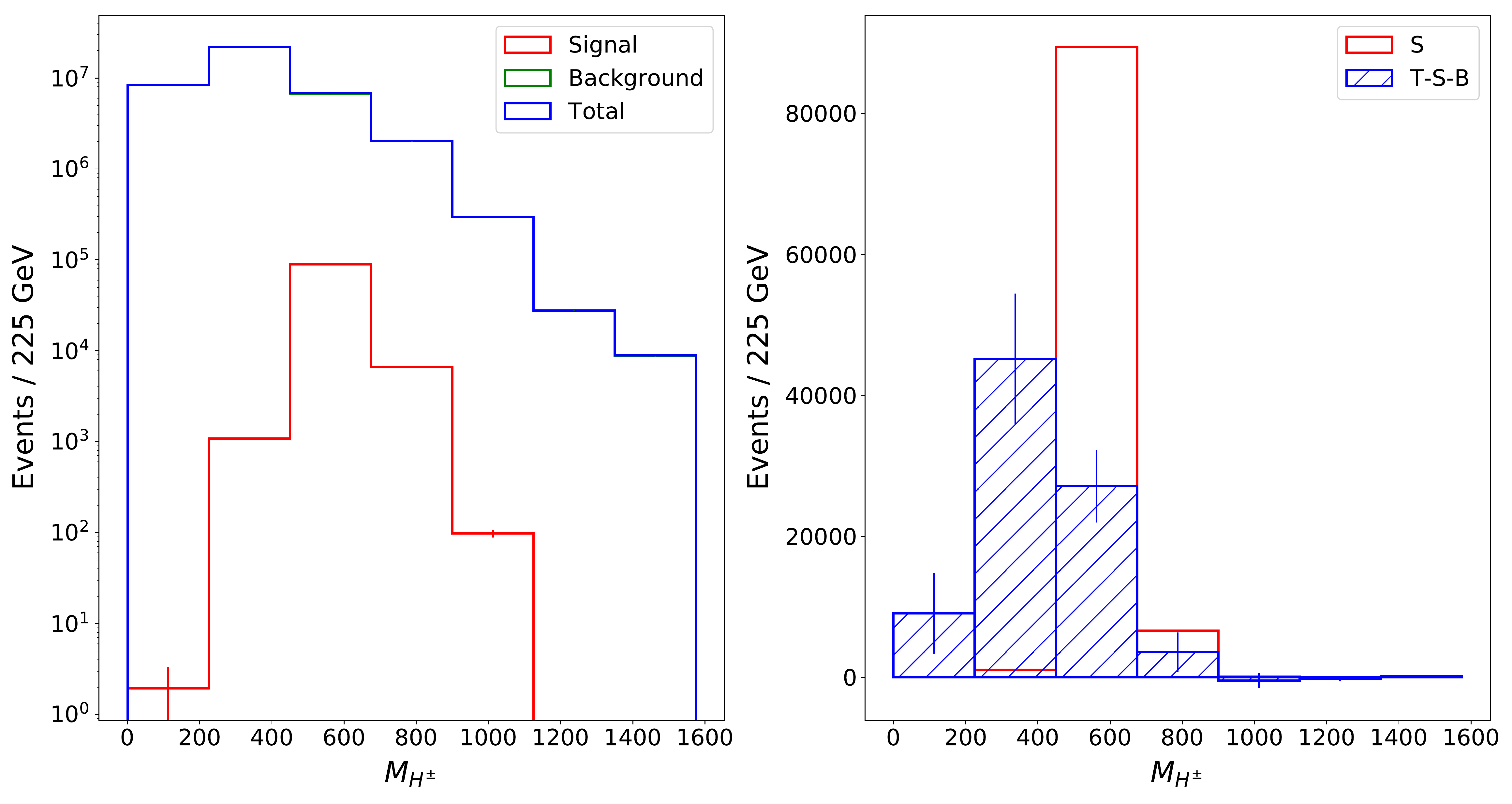}
	\caption{\label{fig:hmssm-interference} The charged Higgs invariant mass distribution  of the signal, background and total samples (left) and interference and signal (right) at parton level {and without cuts} in the $h$MSSM scenario.}
	\end{figure}

In Fig.~\ref{fig:hmssm-interference}, an exploration of this shape {at parton level before cuts} can be seen in the $t\bar{t}b\bar{b}$ reconstructed invariant mass plane. {This step was undertaken at parton level to achieve the required per-bin statistics to discern the shape of the interference distribution.} {There appears to be a large interference impact in across the whole mass range, interestingly though the largest contributions are below the charged Higgs mass peak. The result of this would likely be a smearing of the charged Higgs mass bump towards lower values in actual data.}

    \begin{table}[!ht]
        \centering
        \begin{tabular}{|lccccc|}
        \hline
        \textbf{Cut} & \textbf{S} &   \textbf{B}  & \textbf{T} & \textbf{I} & $\bf{\Delta I}$ \\\hline
        No cuts:                                 &  9720 &  3923550 &  3941700 &   8429 &  2487 \\\hline
        $N_{\ell} = 1$:                          &  2160 &   904247 &   907925 &   1518 &  1193 \\\hline
        $N_J \geq 5$:                            &  1938 &   624001 &   627534 &   1594 &   992 \\\hline
        $N_{BJ} \geq 2$:                         &  1511 &   404919 &   408054 &   1623 &   799 \\\hline
        $\slashed{E} > 20$ GeV:                  &  1435 &   373648 &   376517 &   1433 &   768 \\\hline
        $\slashed{E} + m_T^W > 60$ GeV:          &  1412 &   364026 &   366898 &   1458 &   758 \\\hline\hline
        \textbf{Cut} & \textbf{S} &   \textbf{B}  & \textbf{T} & \textbf{I} & $\bf{\Delta I}$ \\\hline 
        $N_{BJ} \geq 3$:                         &  826 &    171918 &   173430 &    684 &   521 \\\hline
        $\slashed{E} > 20$ GeV:                  &  785 &    158921 &   160376 &    669 &   501 \\\hline
        $\slashed{E} + m_T^W > 60$ GeV:          &  772 &    154880 &   156314 &    660 &   494 \\\hline
        \end{tabular}
        \caption{\label{tab:hmssmresults} Cut flow results presented in expected event yield with 300 fb$^{-1}$ of luminosity for the hMSSM benchmark with 5,000,000 events for each sample.}
    \end{table}

\subsection{The $m_{h}^{\rm mod+}$ analysis}
\begin{figure}[!ht]
	\centering
	\includegraphics[scale=0.3]{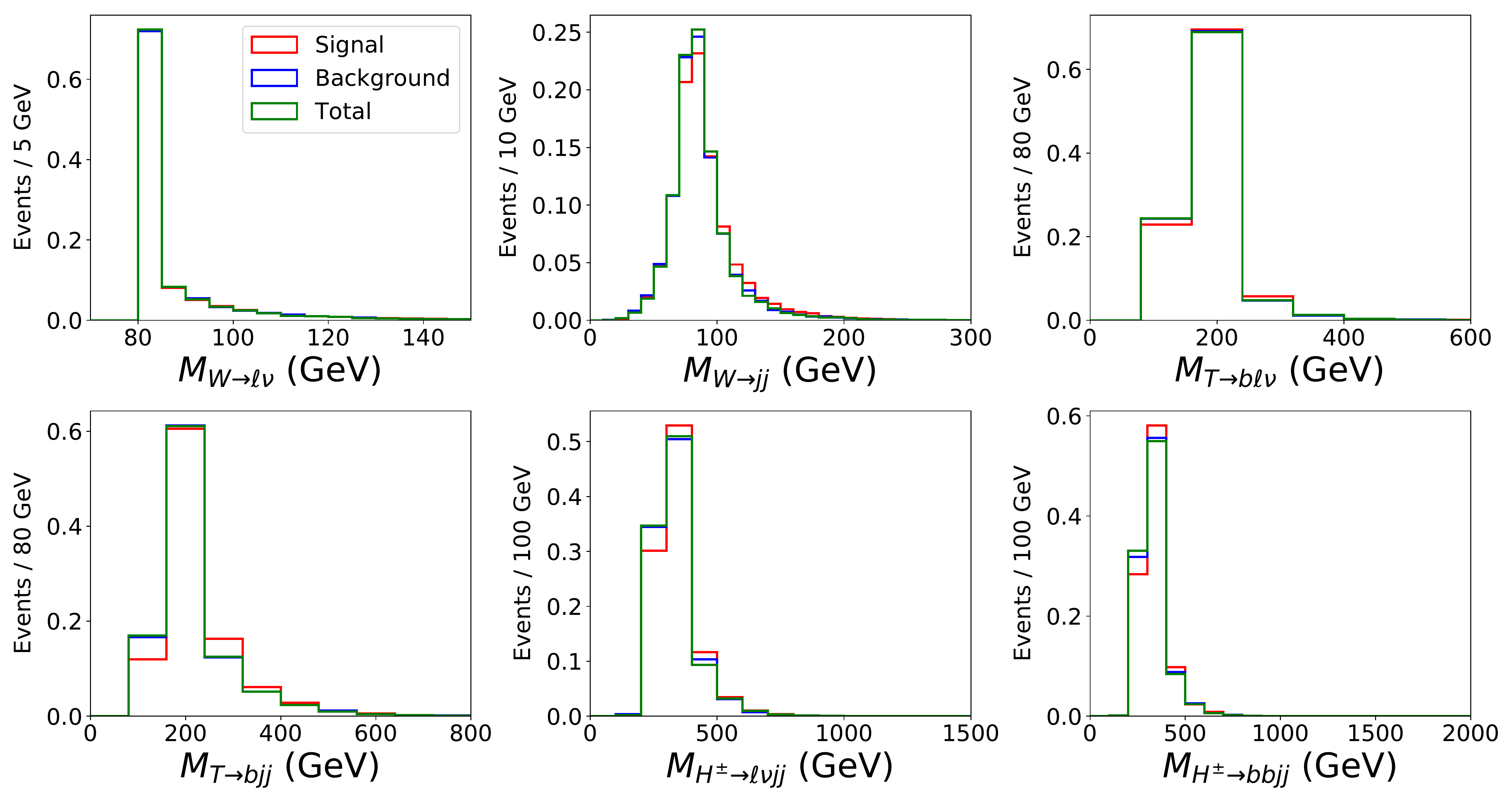}
	\includegraphics[scale=0.3]{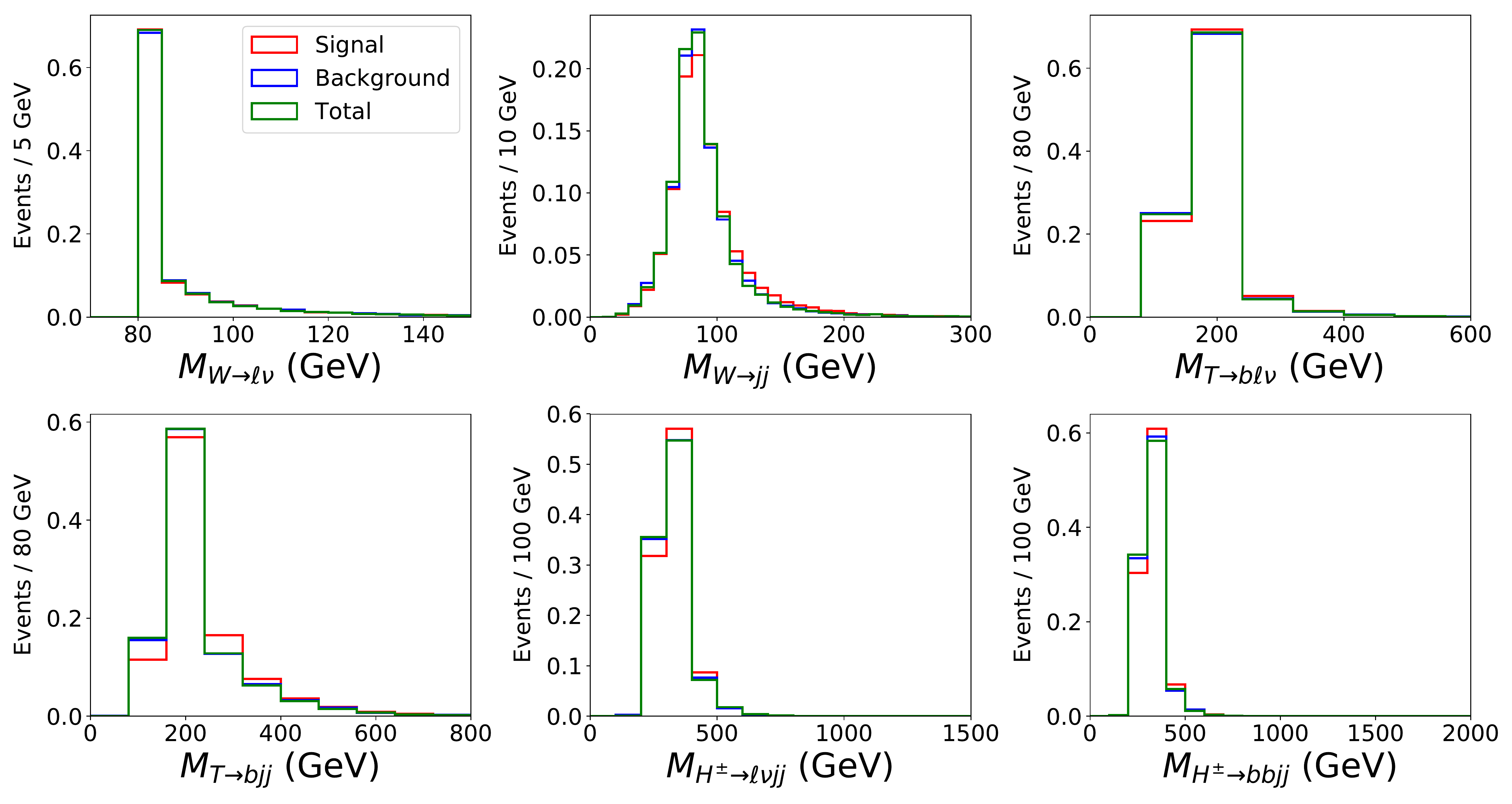}
	\caption{\label{fig:mhmod-recon-variables} Invariant mass distributions for reconstructed particles in the $m_{h}^{\rm mod+}$ benchmark. Top: Model independent. Bottom: Model dependent.}
	\end{figure}

	The reconstruction in this scenario presents far less distinct signal distributions, which can be seen in the charged Higgs invariant mass plots of Fig.~\ref{fig:mhmod-recon-variables}. As the mass difference between the charged Higgs and the sum of the $t$ and $b$-quark masses is far smaller in this scenario, it appears that the reconstruction is performed very similarly for the signal and background. This can be further seen in the lack of  difference between the model-dependent and model-independent reconstructions. Thus, extraction of the signal would be far more difficult in this case.
	
	As the ratio of the charged Higgs mass to charged Higgs width is smaller {in this benchmark} than in the $h$MSSM one, we expect the interference effects to be smaller. However, the interference may become much larger relative to the signal after a cutflow. Tab.~\ref{tab:mhmodresults} displays the cutflow results for this BP and one can see that the pre-cut ratio of interference cross section to signal cross section is $15.2$\%, while the ratio after cuts in the $\geq 2$ $b$-tag region is $26.0$\%.  In this scenario the more strict use of $b$-tagging decreased the ratio of the interference cross section to the signal cross section to $17.4$\%, though it should be noted the uncertainty on the values in the $\geq 3$ $b$-tag region has grown beyond the magnitude of interference and conclusions in this region are tenuous.


	\begin{table}[!ht]
		\centering
		\begin{tabular}{|lccccc|}
		\hline
	    \textbf{Cut} & \textbf{S} &  \textbf{B} & \textbf{T} & \textbf{I} & $\bf{\Delta I}$ \\\hline
	    No cuts:  				 		 & 26561 & 3928500 & 3959100 &  4038 &  1764 \\\hline
        $N_{\ell} = 1$:          		 &  6017 &  903764 &  911292 &  1510 &   846 \\\hline
      	$N_J \geq 5$:            		 &  4964 &  623989 &  629532 &   578 &   703 \\\hline
     	$N_{BJ} \geq 2$:         		 &  3704 &  404776 &  409342 &   862 &   566 \\\hline
    	$\slashed{E} > 20$ GeV:  		 &  3432 &  373464 &  377885 &   989 &   544 \\\hline
 		$\slashed{E} + m_T^W > 60$ GeV:  &  3342 &  363876 &  368087 &   868 &   537 \\\hline\hline
        \textbf{Cut} & \textbf{S} &  \textbf{B} & \textbf{T} & \textbf{I} & $\bf{\Delta I}$ \\\hline
        $N_{BJ} \geq 3$: 				 &  1894 &  171654 &  173822 &   273 &   369  \\\hline
        $\slashed{E} > 20$ GeV:  		 &  1757 &  158581 &  160686 &   347 &   354  \\\hline
        $\slashed{E} + m_T^W > 60$ GeV:  &  1712 &  154576 &  156587 &   298 &   350  \\\hline
		\end{tabular}
		\caption{\label{tab:mhmodresults} Cut flow results presented in expected event yield with 300 fb$^{-1}$ of luminosity for the $m_{h}^{\rm mod+}$ benchmark with 10,000,000 events for each sample.}
	\end{table}

	\begin{figure}[!ht]
	\centering
	\includegraphics[scale=0.3]{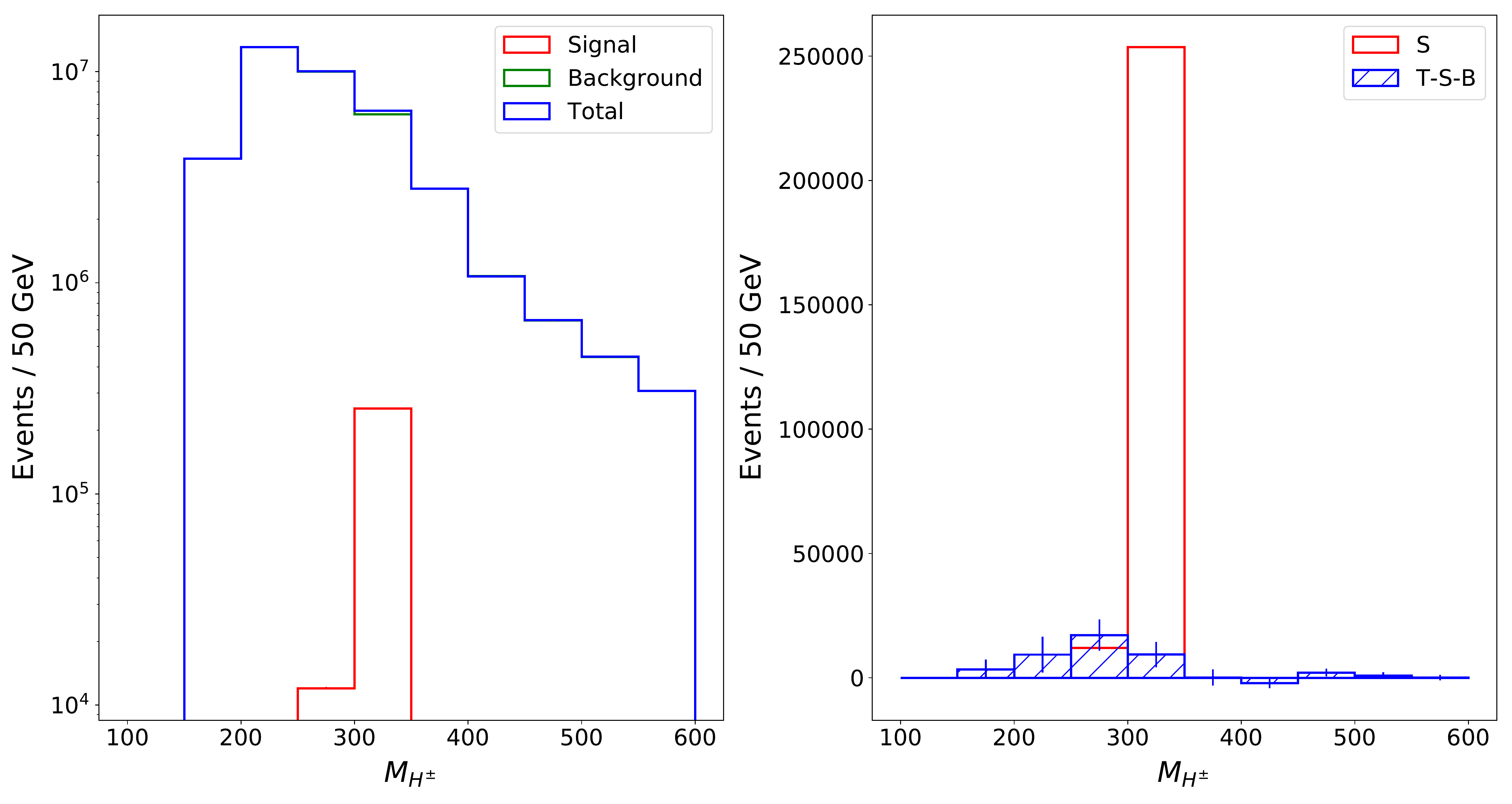}
	\caption{\label{fig:mhmod-interference} The charged Higgs invariant mass distribution  of the signal, background and total samples (left) and interference and signal (right) at parton level {and without cuts} in the $m_{h}^{\rm mod+}$ scenario.}
	\end{figure}
	
	The interference shape of the reconstructed invariant mass of $t\bar{t}b\bar{b}$ at parton level and before cuts in this scenario can be found in Fig~\ref{fig:mhmod-interference}. {In this scenario the signal invariant mass distribution peaks narrowly over the $300$ GeV bin, however, interestingly the interference distribution appears to be a widely spread spectrum across the range $100-350$ GeV. These small, but non-negligible, interference contributions would likely lead to a widening of an otherwise sharp signal bump in data, which further motivates the necessity of the signal shape analysis at detector level to discover whether this effect would be dominated by the smearing effect of the detector on the signal peak.}

\section{Conclusions}
By borrowing the MSSM as a theoretical template that contains charged Higgs bosons, we have shown how experimental searches for these states cannot be made immune from large interference effects between signal and background whenever they have a large mass and a width on the order of one percent of the mass and upwards. We have illustrated this for the case of the $H^+\to t\bar b$ decay channel, which is onset dominantly by $gg\to b\bar t H^+$ production. In this case, the (irreducible) background intervening in such interference effects is $pp\to t\bar t b\bar b$, which can see both QCD and EW interactions. This study's goal was to show that signal and background are wrongly treated as separate in current LHC approaches. 

In order to realistically assess the above phenomenon, we have decayed the $t\bar t$ pair semi-leptonically and carried out a full parton shower, hadronisation and detector analysis. In doing so, we have first prepared the MSSM parameter space regions amenable to phenomenological investigation by enforcing both  theoretical (i.e., unitarity, perturbativity, vacuum stability, triviality) and experimental (i.e., from flavour physics, void and successful Higgs boson searches at the Tevatron and LHC, EW precisions observables from LEP and SLC) constraints, assuming two benchmark configurations of the MSSM, the so-called  $h$MSSM and $m_{h}^{\rm mod+}$ scenarios. 

After performing a sophisticated MC simulation, allowing for both model-independent and model-dependent selections, we have seen that such interference effects can be very large, even of ${\cal O}(100\%)$, both before and after $H^\pm$ detection cuts are enforced. This appears to be the case for the masses tested, approximately 300 and 630 GeV, in the MSSM scenarios adopted, though interference effects will manifest themselves at different LHC stages, depending on the overall cross sections, which vary significantly from one benchmark to another. Furthermore, the shapes of the  signal and its interference (with the aforementioned irreducible background) appear to be different which would mean that it is not actually possible to proceed to a rescaling of the event yields due solely to the signal. In turn, in experimental analyses, one should  account for such interference effects at the event generation level. We have proven this to be the case for a standard cut flow, while deferring the study of similar effects in the case of a machine learning framework to a future publication.




Before closing, though, we highlight the fact that the aforementioned results have been obtained in presence of a statistical error that (in some, but not all, cases) competes with the size of the interferences themselves, owing to a limitation of our computing resources. Nonetheless, in all cases we were able to obtain rather stable cutflows, wherein the increase in statistics essentially reduced the errors as expected without significantly altering the central values of our predictions. On the basis of this, we advocate future experimental analyses with larger data samples to establish the exact extent of such interference effects.

\section*{Acknowledgements}
AA, DA, RB, HH, SM and RS are partially supported by the H2020-MSCA-RISE-2014 grant no. 645722 (NonMinimalHiggs). SM is further supported  through the NExT Institute and the STFC Consolidated Grant ST/L000296/1.
DA and RS are further supported in part by the CERN fund grant CERN/FIS-PAR/0002/2017, by an FCT grant PTDC/FIS-PAR/31000/2017, by the CFTC-UL projects UIDB/00618/2020 and UIDP/00618/2020, 
and by the HARMONIA project under contract UMO-2015/18/M/ST2/00518. RP is supported by the University of Adelaide and the Australian Research Council through the ARC Center of Excellence for Particle Physics (CoEPP) at the Terascale (grant no. CE110001004). This work is also supported by the Moroccan Ministry of Higher Education and Scientific Research MESRSFC and CNRST,  Project PPR/2015/6. We thank Pietro Slavich and Bill Murray for discussions.

\end{document}